\documentclass[aps,prd,floatfix,superscriptaddress,amsmath,amssymb, preprint]{revtex4-2}
\usepackage{amsfonts}
\usepackage{graphicx,color}
\usepackage{times}
\usepackage{amsmath} 

\usepackage{booktabs}

\usepackage{multirow} 

\usepackage{tabularx} 

\usepackage{makecell}

\usepackage{enumitem}

\usepackage{xcolor} 
\usepackage{hyperref} 
\usepackage{float}
\usepackage{hyperref} 

\begin{document}
\title{Forecast Constraints on Bouncing Cosmology from High-Frequency Gravitational Waves Using Superconducting LC Circuits and Resonant Cavities}

\author{Changhong Li}
\email{changhongli@ynu.edu.cn}
\affiliation{Department of Astronomy,  Key Laboratory of Astroparticle Physics of Yunnan Province, School of Physics and Astronomy,  Yunnan University, No.2 Cuihu North Road, Kunming, China 650091}

\begin{abstract}
We exploit forecast sensitivities to high‐frequency gravitational waves (HFGWs) from superconducting LC circuits, traditional resonant cavity and superconducting radio-frequency (SRF) cavities with electromagnetic and mechanical modes to derive the first projections of the bounce energy scale within the generic bouncing‐cosmology framework over the frequency window $1\,$kHz $\lesssim f\lesssim10\,$GHz. In comparison with existing astrophysical limits—spanning $10^{-17}\,$Hz $\lesssim f\lesssim1\,$kHz and based on Planck/BICEP, PTA, and aLIGO+LISA—our HFGW forecasts yield substantially tighter constraints across a broad region of parameter space. This work unifies constraints from cosmological observations and quantum‐measurement experiments, providing comprehensive coverage of the early‐Universe gravitational‐wave spectrum from $10^{-17}\,\mathrm{Hz}$ to $10\,\mathrm{GHz}$ and thereby probing the cosmic initial non‐singularity at ultra‐high energy scales.
\end{abstract}

\date{\today}

\pacs{}
\maketitle

\newpage
  
\section{Introduction}
Bouncing cosmology~\cite{Novello:2008ra,Brandenberger:2016vhg,Nojiri:2017ncd,Odintsov:2023weg}, in which the Universe undergoes a non‐singular transition from contraction to expansion~\cite{Khoury:2001wf,Gasperini:2002bn,Creminelli:2006xe,Peter:2006hx,Cai:2007qw,Cai:2008qw,Saidov:2010wx,Li:2011nj,Cai:2011tc,Easson:2011zy,Bhattacharya:2013ut,Qiu:2015nha,Barrow:2017yqt,deHaro:2017yll,Boruah:2018pvq,Nojiri:2019yzg,Silva:2015qna,Silva:2020bnn,Silva:2023ieb,Tukhashvili:2023itb,Garcia-Saenz:2024ogr,Li:2024rgq,Nayeri:2005ck,Brandenberger:2006xi,Fischler:1998st,Cai:2009rd,Cheung:2014nxi,Li:2014cba,Li:2015egy}, provides a compelling alternative to standard inflationary scenarios~\cite{Guth:1980zm,Starobinsky:1980te,Sato:1980yn,Linde:1981mu,Albrecht:1982wi,Mukhanov:1990me} for resolving the Universe’s initial singularity~\cite{Borde:1993xh,Borde:2001nh}. Akin to inflation, bouncing cosmology also addresses the horizon and flatness problems~\cite{Ijjas:2018qbo} and yields a nearly scale‐invariant primordial curvature perturbation compatible with current CMB anisotropy measurements~\cite{WMAP:2010qai,Planck:2015fie,Planck:2018vyg} (e.g.,~\cite{Li:2013bha}). Moreover, late‐time processes such as leptogenesis~\cite{Barrie:2021orn} and dark matter production~\cite{Li:2014era,Li:2020nah} have been extensively studied in this framework.

Among observational probes, stochastic gravitational waves—recently indicated by pulsar timing arrays~\cite{NANOGrav:2023gor,EPTA:2023fyk,Reardon:2023gzh,Antoniadis:2022pcn,Xu:2023wog,NANOGrav:2023hvm,EPTA:2023xxk,Figueroa:2023zhu,Bian:2023dnv,Ellis:2023oxs,Lai:2025xov}—offer a decisive window into early‐Universe dynamics, including both inflationary and bouncing scenarios~\cite{Caprini:2018mtu} (e.g.,~\cite{Zhao:2013bba,Guzzetti:2016mkm,Cai:2016hea,Vagnozzi:2020gtf,Benetti:2021uea,Vagnozzi:2023lwo,Zhu:2023lbf,Papanikolaou:2024fzf,Li:2024oru}). In particular, for bouncing cosmology, the stochastic background induced by primordial gravitational waves (PGWs) may unveil the fine‐structure of the initial non‐singular phase (e.g.,~\cite{Li:2024oru,Qiu:2024sdd,Lai:2025efh}).

However, compared to inflation, the cosmic evolution of PGWs in a bouncing Universe is more intricate due to the non‐singular phase: PGWs exit and reenter the horizon twice, whereas in inflation they do so only once~\cite{Cheung:2016vze}. Consequently, deriving a complete expression for the PGW spectrum over the full evolution of a generic bouncing model has been challenging~\cite{Boyle:2004gv,Li:2013bha,Cai:2014bea,Li:2024oru}. Recently, by developing an orthogonal algebra for the power‐law indices of growing and decaying modes, an analytical matrix representation of the PGW spectrum over the entire evolution of a generic bouncing Universe was obtained (Eq.~\eqref{eq:matrixPGW}) in Ref.~\cite{Li:2024dce}. This formalism enables us to constrain the pivot cosmological parameters of a bouncing model via GW observations. In particular, because the bouncing phase is explicitly included, the PGW spectrum depends on the bounce energy scale \(\rho_{s\downarrow}^{1/4}\) (Eq.~\eqref{eq:rhosdd}), which can now be constrained by current and future GW searches.

Building on this framework, Ref.~\cite{Li:2025ilc} constructs a concrete bouncing‐cosmology model in which the bounce is assumed to be rapid and symmetric, with the post‐bounce Universe reheating directly into a standard radiation‐dominated phase. By inserting these assumptions into the general matrix representation of the primordial‐GW spectrum derived in Ref.~\cite{Li:2024dce}, the stochastic gravitational‐wave background (SGWB) of this specific bounce model is obtained. Using the sensitivities of current GW observatories—Planck/BICEP, PTA and LIGO/LISA—and projected sensitivities of forthcoming experiments (CMB‐S4, IPTA (design), SKA, DECIGO, BBO, LISA, TianQin, Taiji, aLIGO+Virgo+KAGRA (design), Cosmic Explorer and the Einstein Telescope; see Refs.~\cite{Schmitz:2020syl,Annis:2022xgg,Bi:2023tib} for details), Ref.~\cite{Li:2025ilc} reports the first systematic GW limit on the bounce energy scale \(\rho_{s\downarrow}^{1/4}\) (which corresponds to the quasi‐highest temperature of the bounce) and the associated bounce scale factor \(a_{s\downarrow}\) (the quasi‐minimal size of the Universe). These results exclude a significant region of parameter space for \(\rho_{s\downarrow}^{1/4}\), particularly when \(-\tfrac{1}{3}<w_1<0\) (see Fig.~\ref{fig:LCSRF}), where \(w_1\) denotes the contraction‐phase equation‐of‐state.

However, we note that these astrophysical and laser GW detectors primarily probe the low‐frequency band \(10^{-17}\,\mathrm{Hz}\lesssim f\lesssim1\,\mathrm{kHz}\) and cannot constrain bouncing cosmology at higher frequencies (e.g.\ \(1\,\mathrm{kHz}<f<10\,\mathrm{GHz}\)). As a result, the bounds on \(\rho_{s\downarrow}^{1/4}\) remain relatively weak for models with \(w_1>0\).

In this study, we exploit forecast sensitivities to high‐frequency gravitational waves (HFGWs) from superconducting LC circuits, conventional resonant cavities, and superconducting radio‐frequency (SRF) cavities with both electromagnetic and mechanical modes to derive the first projections of \(\rho_{s\downarrow}^{1/4}\) over the frequency window \(1\,\mathrm{kHz}\lesssim f\lesssim10\,\mathrm{GHz}\). These LC circuits and cavities were originally designed to search for axions and dark photons~\cite{Sikivie:1983ip,Sikivie:1985yu,Sikivie:2013laa,Chaudhuri:2014dla,Kahn:2016aff}. Thanks to the inverse Gertsenshtein effect (graviton–photon conversion in a magnetic field) and mechanical resonance~\cite{Berlin:2023grv}, these setups can also be used to probe HFGWs over \(1\,\mathrm{kHz}\lesssim f\lesssim10\,\mathrm{GHz}\) (for a comprehensive review, see~\cite{Chen:2023ryb}). Incorporating these HFGW constraints yields substantially tighter bounds on \(\rho_{s\downarrow}^{1/4}\) for \(w_1>0\). For instance, whereas Planck/BICEP provides the strongest constraint for \(-\tfrac{1}{3}<w_1<0\), the SRF EM channel improves the bound on \(\rho_{s\downarrow}^{1/4}\) by over twenty orders of magnitude in the regime \(w_1\gg0\). Our results highlight the potential of circuits and cavities to explore the very early Universe and underscore the necessity of unifying constraints from cosmological observations and quantum‐measurement experiments. This unified approach offers comprehensive coverage of the early‐Universe gravitational‐wave spectrum from \(10^{-17}\,\mathrm{Hz}\) to \(10\,\mathrm{GHz}\), probing new physics at unprecedented energy scales.

\section{The SGWB Spectrum in Bouncing Universe}

The general matrix representation of primordial gravitational wave spectrum in a generic bouncing universe is given by~\cite{Li:2024dce} 
\begin{equation}\label{eq:matrixPGW}
    \mathcal{P}_h(\eta_k) \equiv \frac{k^3(|h_{k+}^{[4]}|^2 + |h_{k \times}^{[4]}|^2)}{2\pi^2}
    = \frac{(k\eta_k)^3}{2\pi^2} \frac{\pi}{\nu_4^2} \frac{[H(\eta_k)]^2}{m_\mathrm{pl}^2} N_{22}(\{\Tilde{\nu}_i\}, \{\eta_{i\downarrow/\uparrow}\})~,
\end{equation}
Here $h_{k+/\times}^{[i]}$ are the Fourier modes of metric tensor perturbation $h_{ij}^{TT}$ with commoving wavevector $k$ in the transverse–traceless gauge. $i=1,2,3,4$ denotes the phase I (collapsing contraction), phase II (bouncing contraction), phase III (bouncing expansion) and phase IV (post-bounce expansion) respectively, $\eta_i$ is conformal time and $\eta_k=k^{-1}$ is the horizon re-entering moment in phase IV. $H(\eta_k)$ is Hubble parameter at $\eta=\eta_k$, and $m_\mathrm{pl}\equiv \sqrt{8\pi G}^{-1}$ is reduced Planck mass. The power-law index $\nu_i$  and the reduced power-law index $\tilde{\nu}_i\equiv |\nu_i-\tfrac{1}{2}|$ of scalar factor for each phase, $a(\eta)=a_i|\eta|^{\nu_i}$, are, respectively, given by 
\begin{equation}\label{eq:nutnuw}
    \nu_i=\frac{2}{3w_i+1} \quad\mathrm{and}\quad \Tilde{\nu}_i=\left|\frac{3(1-w_i)}{2(3w_i+1)}\right|~,
\end{equation}
where $w_i$ is background equation of state (EoS), assumed to be constant, for each phase.

The kernel term $N_{22}(\{\Tilde{\nu}_i\}, \{\eta_{i\downarrow/\uparrow}\})$ is the $_{22}$-component of the propagation matrix for PGW spectrum in bouncing universe, $N(\{\Tilde{\nu}_i\}, \{\eta_{i\downarrow/\uparrow}\})$, 
\begin{equation}\label{eq:Npgwsm}
    N(\{\Tilde{\nu}_i\}, \{\eta_{i\downarrow/\uparrow}\}) \equiv X^\dagger X~,\quad 
\end{equation}
and $X$ is the propagation matrix for PGW amplitude, 
\begin{equation}\label{eq:Xpgwam}
    X \equiv T_4^{-1} M_{3\downarrow} T_3 M_{2\uparrow} T_2^{-1} M_{1\downarrow} T_1~,
\end{equation}
which consists of the transformation matrices $T_i$ (or their inverses $T_i^{-1}$) in each phase, 
\begin{equation}
    T_i = 
    \begin{pmatrix}
        \alpha_i & \alpha_i^\ast \\
        \beta_i & \beta_i^\ast
    \end{pmatrix}~,
\end{equation}
and the matching matrices $M_{i\uparrow/\downarrow}$ between boundaries~\cite{Li:2024dce}: 

\begin{equation}
 M_{1\downarrow}=
    \begin{cases}
        \begin{pmatrix}
        0 & -\frac{\Tilde{\nu}_1}{\Tilde{\nu}_2} \left(k \eta_{1\downarrow}\right)^{-\left(\Tilde{\nu}_1 + \Tilde{\nu}_2\right)} \\
        \left(k \eta_{1\downarrow}\right)^{\Tilde{\nu}_1 + \Tilde{\nu}_2} & \left(1 + \frac{\Tilde{\nu}_1}{\Tilde{\nu}_2}\right) \left(k \eta_{1\downarrow}\right)^{-\left(\Tilde{\nu}_1 - \Tilde{\nu}_2\right)}
    \end{pmatrix}~,\quad \nu_1 > \tfrac{1}{2}~,\\
    \begin{pmatrix}
        \frac{\Tilde{\nu}_1}{\Tilde{\nu}_2} \left(k \eta_{1\downarrow}\right)^{\Tilde{\nu}_1 - \Tilde{\nu}_2} & 0 \\
        \left(1 - \frac{\Tilde{\nu}_1}{\Tilde{\nu}_2}\right) \left(k \eta_{1\downarrow}\right)^{\Tilde{\nu}_1 + \Tilde{\nu}_2} & \left(k \eta_{1\downarrow}\right)^{-\left(\Tilde{\nu}_1 - \Tilde{\nu}_2\right)}
    \end{pmatrix}~,\quad \nu_1 \le \tfrac{1}{2}~,
    \end{cases}
\end{equation}

\begin{equation}\label{eq:m2exp}
    M_{2\uparrow}
    = \begin{pmatrix}
        e^{-i(\Tilde{\nu}_2 - \Tilde{\nu}_3)\pi/2} & 0 \\
        0 & e^{i(\Tilde{\nu}_2 - \Tilde{\nu}_3)\pi/2}
    \end{pmatrix}~, \quad  \nu_2 < 0 \quad \mathrm{and}\quad  \nu_3 < 0~,  
\end{equation}

\begin{equation}
    M_{3\downarrow} =
    \begin{cases}
        \begin{pmatrix}
        \left(1 + \frac{\Tilde{\nu}_3}{\Tilde{\nu}_4}\right) \left(k \eta_{3\downarrow}\right)^{\Tilde{\nu}_3 - \Tilde{\nu}_4} & \left(k \eta_{3\downarrow}\right)^{-\left(\Tilde{\nu}_3 + \Tilde{\nu}_4\right)} \\
        -\frac{\Tilde{\nu}_3}{\Tilde{\nu}_4} \left(k \eta_{3\downarrow}\right)^{\Tilde{\nu}_3 + \Tilde{\nu}_4} & 0
    \end{pmatrix}~,\quad \nu_4 > \tfrac{1}{2}~.\\
    \begin{pmatrix}
        \frac{\Tilde{\nu}_3}{\Tilde{\nu}_4} \left(k \eta_{3\downarrow}\right)^{\Tilde{\nu}_3 - \Tilde{\nu}_4} & 0 \\
        \left(1 - \frac{\Tilde{\nu}_3}{\Tilde{\nu}_4}\right) \left(k \eta_{3\downarrow}\right)^{\Tilde{\nu}_3 + \Tilde{\nu}_4} & \left(k \eta_{3\downarrow}\right)^{-\left(\Tilde{\nu}_3 - \Tilde{\nu}_4\right)}
    \end{pmatrix}~,\quad \nu_4 \le \tfrac{1}{2}~,
    \end{cases}
\end{equation}
where $\nu_2 < 0$ and $\nu_3 < 0~$ in Eq.~\eqref{eq:m2exp} is the condition for achieving bounce. \begin{equation}
    \alpha_i \equiv 2^{-\Tilde{\nu}_i} \left[ \frac{-i e^{i\pi \Tilde{\nu}_i}}{\sin(\pi \Tilde{\nu}_i) \Gamma(\Tilde{\nu}_i + 1)} \right]~, \quad 
    \beta_i \equiv -i \left[ \frac{\Gamma(\Tilde{\nu}_i)}{\pi} 2^{\Tilde{\nu}_i} \right]~,
\end{equation}
and $\Gamma(z) = \int_0^\infty t^{z-1} e^{-t} \, dt$ is the Gamma function. 

Considering a realistic realization of bouncing cosmology by taking following three physical assumptions: (1) after bouncing, the Universe reheating into the standard post-reheating radiation era ($w_4=\tfrac{1}{3}$); (2) the bouncing process can be achieved, ($ w_2 = w_3 = -\infty $, e.g., Quintom matter~\cite{Cai:2009zp}), note that, in model-building, $w_2=w_3<-5$ reproduces the same result with sufficient accuracy; and (3) we assume a symmetric bounce, $ \eta_{s\downarrow} \equiv \eta_{1\downarrow} = \eta_{3\downarrow} $. Mathematically, this bouncing universe model can be expressed as
\begin{equation}\label{eq:wncon}
    \begin{cases}
        \{w_i\}=(w_1, w_2, w_3, w_4) = \left(w_1, -\infty, -\infty, \tfrac{1}{3}\right)~,\\
         \{\eta_{i\downarrow/\uparrow}\}=(\eta_{1\downarrow},\eta_{2\uparrow},\eta_{3\downarrow})=(\eta_{s\downarrow},\infty,\eta_{s\downarrow})~,
    \end{cases}
\end{equation}
where $\eta_{2\uparrow}\rightarrow \infty$ is for the bouncing point ($\dot{a}=0$). Substituting Eq.~\eqref{eq:wncon} into Eq.~\eqref{eq:nutnuw} and Eq.~\eqref{eq:Npgwsm}, we obtain~\cite{Li:2025ilc}

\begin{equation} \label{eq:Npowlaw}
    N_{22}(w_1) = C(w_1) \frac{1}{(k\eta_{s\downarrow})^{n(w_1)}}~,
\end{equation}
with 
\begin{equation}\label{eq:Cw1}
    C(w_1) = 
\begin{cases} 
    \pi^{-1}4^{-1+\frac{3(1-w_1)}{2(3w_1+1)}}\Gamma^2\left(\frac{3(1-w_1)}{2(3w_1+1)}\right)\left[\frac{2(3w_1-1)}{3w_1+1}\right]^2, & \text{if } -\tfrac{1}{3} \leq w_1 < 1  \\
    \pi^{-1}4^{-1-\frac{3(1-w_1)}{2(3w_1+1)}}\Gamma^2\left(-\frac{3(1-w_1)}{2(3w_1+1)}\right), & \text{if } w_1\ge1 
\end{cases},
\end{equation}
and 
\begin{equation}\label{eq:nw1}
    n(w_1) = 
\begin{cases} 
    1 - \frac{3(w_1-1)}{(3w_1+1)}, & \text{if } -\tfrac{1}{3} \leq w_1 < 1 \\
    1 + \frac{3(w_1-1)}{(3w_1+1)}, & \text{if } w_1\ge1 
\end{cases},
\end{equation}
where we have used the deep bouncing limit $k\eta_{s\downarrow}\ll 1$ for $k$ of interest to simplify the expression and obtain this leading order term of $ N_{22}(\{\Tilde{\nu}_i\}, \{\eta_{i\downarrow/\uparrow}\})$.

Substituting Eq.~\eqref{eq:Npowlaw} into Eq.~\eqref{eq:matrixPGW}, and using the expression of SGWB spectrum,
\begin{equation}\label{eq:sgwbspectrum}
    \Omega_\mathrm{GW}(f)h^2 = \frac{1}{24}\Omega_{\gamma 0} h^2 \cdot \mathcal{P}_h(f) \mathcal{T}_{\mathrm{eq}}(f),
\end{equation}
we obtain 
\begin{equation}\label{eq:OGWrhos}
    \Omega_\mathrm{GW}(f)h^2 
    =\frac{h^2}{24} \left(\frac{f_{H_0}}{f_{m_\mathrm{pl}}}\right)^2  \cdot \frac{C(w_1)}{(2\pi)^{-n_T(w_1)-1}} \left(\frac{f}{f_{H_0}}\right)^{n_T(w_1)} \left[\frac{\rho_{s\downarrow}^{1/4}}{\left(\rho_{c0}/ \Omega_{\gamma 0}\right)^{1/4}}\right]^{4-n_T(w_1)}  \mathcal{T}_{\mathrm{eq}}(f),
\end{equation}
where $f=2\pi k/a_0$ is the frequency observed today, $f_{H0}=2.2\times 10^{-18} ~\mathrm{Hz}$ corresponds to $H_0$ and $f_{m_\mathrm{pl}}=3.7\times 10^{42} ~\mathrm{Hz}$ corresponds to the reduced Planck mass $m_\mathrm{pl}$ in natural units ($\hbar=1$ and $c=1$). For the modes of interest in this work ($f \gtrsim f_\mathrm{eq}$), the transfer function $\mathcal{T}_{\mathrm{eq}}(f) \equiv \left[1 + \frac{9}{32} \left( f_\mathrm{eq}/f \right)^2 \right] \simeq 1$ with $ f_\mathrm{eq} = 2.01 \times 10^{-17}~\mathrm{Hz} $. $ \Omega_{\gamma 0} h^2 = 2.474 \times 10^{-5} $ is the energy density fraction of radiation today with $ h = 0.677 $, and $\rho_{c0} = 3H_0^2 m_\mathrm{pl}^2 $ is the the critical energy density today. The energy scale at the onset of post-bounce reheating (bounce energy scale/quasi-highest energy scale), $\rho_{s\downarrow}^{1/4}$, is given by
\begin{equation}\label{eq:rhosdd}
    \rho_{s\downarrow}^{1/4} \equiv \rho (\eta_s)^{1/4}= \left[3H^2(\eta_{s\downarrow}) m_\mathrm{pl}^2\right]^{1/4}~.
\end{equation}
$ n_T $ is the spectral index of the PGWs,
\begin{equation}\label{eq:nt}
    n_T \equiv \frac{d \ln \mathcal{P}_h(\eta_k)}{d \ln k} = 4 - n(w_1)=3-\left|\frac{3(w_1-1)}{(3w_1+1)}\right|~.
\end{equation}
Using Eq.~\eqref{eq:OGWrhos}, the bounce energy scale $\rho_{s\downarrow}^{1/4}$ can be constrained as a function of the contraction equation‐of‐state parameter $w_1$ by existing astrophysical observations and stochastic gravitational‐wave experiments—such as CMB/BICEP, LIGO/Virgo/KAGRA, PTA/SKA and LISA—over the low‐frequency band $10^{-17}\,\mathrm{Hz}\lesssim f\lesssim1\,\mathrm{kHz}$~\cite{Li:2025ilc,Lai:2025efh}. In this work, we exploit forecast sensitivities to high‐frequency gravitational waves (HFGWs) from resonant cavities and superconducting circuits to extend these constraints on $\rho_{s\downarrow}^{1/4}$ into the high‐frequency window $1\,\mathrm{kHz}\lesssim f\lesssim 10\,\mathrm{GHz}\,$.

\section{The Sensitivity of High-frequency Gravitational Waves Probed in Resonant Cavities and Superconducting Circuits}

The gravitational wave signal probed by resonant cavities and superconducting circuits at a given frequency can be approximated as a monochromatic plane wave in the transverse–traceless gauge:
\begin{equation}
h_{ij}^{\mathrm{TT}}(t,x)
= h_{0}\,H_{ij}^{\mathrm{TT}}\,
e^{i(\omega_h t - k x)} \,,
\end{equation}
where \(h_{0}\) is the dimensionless strain amplitude, \(H_{ij}^{\mathrm{TT}}\) is the normalized polarization tensor satisfying \(H_{ij}^{\mathrm{TT}}H_{ij}^{\mathrm{TT}} = 2\), and \(\omega_h=2\pi f\) is the angular frequency.

The corresponding energy density is
\begin{equation}\label{eq:rhogw}
\rho_\mathrm{ GW}
= \frac{1}{32\pi G}\bigl\langle \dot h_{ij}^\mathrm{ TT}\,\dot h_{ij}^\mathrm{ TT}\bigr\rangle
= \frac{\omega_h^2\,h_{0}^2}{16\pi G}\,.
\end{equation}
and the SGWB spectrum to be probed at \(f=f_0\) takes 
\begin{equation}\label{eq:Omehzero}
\Omega_\mathrm{ GW}(f)
= \frac{1}{\rho_{c}}\frac{d\rho_\mathrm{ GW}}{d\ln f}
= \frac{\rho_\mathrm{ GW}}{\rho_{c}}
= \frac{2\pi^2\,f^2\,h_{0}^2}{3\,H_{0}^2}\,,
\end{equation}
where $\rho_{c} = \frac{3H_{0}^2}{8\pi G}$ is the critical energy density today, and $\frac{d\rho_\mathrm{ GW}}{d\ln f}\Big|_{f_0}=\rho_\mathrm{ GW}$ is used for a monochromatic signal (If we treat $\rho_\mathrm{GW}$ as smooth function rather than a monochromatic signal, it should be different by a factor $2$, $\frac{d\rho_\mathrm{ GW}}{d\ln f}=2\rho_\mathrm{ GW}$). 

Substituting Eq.~\eqref{eq:Omehzero} into Eq.~\eqref{eq:OGWrhos}, we obtain
\begin{equation}\label{eq:hzerorhos} 
    h_{0}^2=\frac{1}{4} \left(\frac{f_{H_0}}{f_{m_\mathrm{pl}}}\right)^2  \cdot \frac{C(w_1)}{(2\pi)^{-n_T(w_1)+1}} \left(\frac{f}{f_{H_0}}\right)^{n_T(w_1)-2} \left[\frac{\rho_{s\downarrow}^{1/4}}{\left(\rho_{c0}/ \Omega_{\gamma 0}\right)^{1/4}}\right]^{4-n_T(w_1)},
\end{equation}
where we have used $\mathcal{T}_{\mathrm{eq}}(f) = 1$ for HFGW. 

The sensitivity reach of $h_0$ in resonant cavities and superconducting circuits can be estimated by requiring the signal-to-ratio (SNR) to be unity ($\mathrm{SNR}^2=1$)~\cite{Chen:2023ryb},
\begin{equation}
    \mathrm{SNR}^2=\frac{t_\mathrm{int}}{2\pi}\int_0^\infty \left(\frac{S_\mathrm{signal}}{S_\mathrm{noise}}\right)^2 d\omega 
\end{equation}
where $S_\mathrm{signal}$ and $S_\mathrm{noise}$ are, respectively, the signal and noise power spectral densities (PSDs), and $t_\mathrm{int}$ is the integration time. As detailed analysis in Ref.~\cite{Chen:2023ryb}, for the {\bf single-mode} resonant detection, the $\mathrm{SNR}^2$ of LC Circuit ($1~\mathrm{kHz}-100~\mathrm{MHz}$), traditional Cavity ($Q_\mathrm{int} \sim 10^4$, $1~\mathrm{GHz}-10~\mathrm{GHz}$)  and superconducting radio-frequency (SRF) cavity with significantly high quality factors ($Q_\mathrm{ int} \gg 10^9$, $1~\mathrm{kHz}-1~\mathrm{GHz}$) for electromagnetic $\mathrm{SRF^{EM}}$ and mechanical resonances $\mathrm{SRF^{mech}}$ for HFGW takes (c.f., Eq.(46) in Ref.~\cite{Chen:2023ryb}),

\begin{equation}\label{eq:single}
   \mathrm{SNR}_\mathrm{single}^2=
    \begin{cases}
        \frac{1}{8\pi} h_0^4 \omega_h^8\eta^4  B_0^4 V^{{14}/{3}}  Q_h Q_{\text{int}} t_e/T,& \mathrm{LC\ Circuit}\\
        \frac{3}{4\pi} h_0^4 \omega_h^{3} \eta^4 B_0^4 V^{{10}/{3}}   Q_{h}Q_{\text{int}}t_e, &\mathrm{Cavity}\\
        \frac{1}{32\pi} h_0^4 \omega_h^6 \eta^4  B_0^4 V^{{14}/{3}}  Q_h Q_\mathrm{ int} t_e \omega_\mathrm{ rf}^2/T, & \mathrm{SRF^{EM}}\\
         \frac{h_0^4 \omega_h^6 \vert \eta_p^t\,\eta_p^h\,L_p(\omega_h)\vert^4
 B_0^4 V^2 Q_h Q_{\text{int}} t_e \omega_{\text{rf}}^2}{16\pi T N_M} 
  \times \mathrm{ min}\left(1, \frac{Q_\mathrm{ int}\omega_h}{T N_M}\right), & \mathrm{SRF^{mech}}
    \end{cases}~,
\end{equation}
with the benchmark parameters~\cite{Berlin:2023grv} (c.f., section 4 in Ref.~\cite{Chen:2023ryb}),
\begin{equation}
    \begin{cases}
        B_0 = 4\, \mathrm{T}, Q_\mathrm{int} = 10^6, T = 0.01\,\mathrm{K}, V = 1\,  \mathrm{m}^3; &\mathrm{LC\ circuit} \\
         B_0 = 4\, \mathrm{T}, Q_\mathrm{int}=10^4, T = 0.01\,\mathrm{K}, V = 1\,  \mathrm{m}^3;&
\mathrm{Cavity}\\
                B_0 = 0.2\, \mathrm{T}, Q_{\mathrm{int}} = 10^{12}, T = 1.8\,\mathrm{K}, V = 1\,  \mathrm{m}^3, \omega_\mathrm{rf} \approx \omega_0 = 2\pi\,\mathrm{GHz}, & \mathrm{SRF^{EM/mech}} \\Q_p=10^6, \omega_p=10\,\mathrm{kHz}, M_S=10\,\mathrm{kg}, \eta_p^t=1, \eta_p^h=0.18, &\\
    \end{cases}
\end{equation}
$t_e=10^7\mathrm{s}$ is the e-fold time, $Q_h=10^3$ and $\eta=0.1$ are, respectively, the quality factor and the electromagnetic coupling for GW detection~\cite{Berlin:2021txa, Domcke:2022rgu}, the response function $L_p(\omega_h)$ is given by 
\begin{equation}
    L_p(\omega_h)=(\omega_h^2 - \omega_p^2 + i \omega_h \omega_p/Q_p)^{-1},
\end{equation}
and the dimensionless function $N_M(\omega_h)$ is given by 
\begin{equation}
    N_M(\omega_h)\equiv 1 + Q_{\text{int}} \,\omega_{\text{rf}}\, B_0^2\,\left(4\pi M_S \omega_p T / Q_p\right) \vert\eta_p^t\, L_p(\omega_h) \vert^2\, V^{1/3}/(\pi T M_S^2),
\end{equation}
Here, $B_0$, $Q_{\mathrm{int}}$, $T$ and $V$ denote the background magnetic field, the intrinsic quality factor, the temperature and the resonator volume, respectively; $M_{S}$ is the total mass of the cavity shell; $\eta_{p}^{h}$ and $\eta_{p}^{t}$ are the overlap function and the transition form factor; $\omega_{p}$ and $Q_{p}$ are the resonant frequency and mechanical quality factor of the $p$-th mode; $\omega_{\mathrm{rf}}$ is the eigenfrequency of each mode; and $\omega_{0}$ is the characteristic frequency of the cavity-shell thermal vibrations.

Furthermore, as discussed in Ref.~\cite{Chen:2023ryb}, by connecting $N$ multiple auxiliary modes to the probing sensor, the {\bf single-mode} resonator can be upgraded to a {\bf multi-mode} resonator, which enable broadband detection and achieve a higher sensitivity in large $N$ multi-mode limits: 
\begin{equation}\label{eq:multiple}
   \mathrm{SNR}_\mathrm{multiple}^2=
    \begin{cases}
        \frac{1}{8\pi} h_0^4 \omega_h^9 \eta^4  B_0^4 V^{{14}/{3}}  Q_h Q_{\text{int}}^2 t_e/T^2,& \mathrm{LC\ Circuit}\\
        \frac{1}{2\pi} h_0^4 \omega_h^3 \eta^4  B_0^4 V^{{10}/{3}}  Q_h Q_{\text{int}}^2 t_e, &\mathrm{Cavity}\\
        \frac{3\,\ln{10}}{16\pi} h_0^4 \omega_h^7 \eta^4  B_0^4 V^{{14}/{3}}  Q_h Q_{\text{int}}^2 t_e\omega_{\rm rf}^2/T^2, & \mathrm{SRF^{EM}}\\
         \frac{3\,\ln{10}}{16\pi} \frac{h_0^4 \omega_h^7 \vert \eta_p^t\,\eta_p^h\,L_p(\omega_h)\vert^4 B_0^4 V^{2} Q_h Q_{\text{int}}^2 t_e \omega_{\text{rf}}^2}{T^2 N_M^2}, & \mathrm{SRF^{mech}}
    \end{cases}~.
\end{equation}
Here, $\mathrm{SNR}_\mathrm{multiple}^2$ are computed in the limit, $N\rightarrow \infty$. In practice, $N=5$ is sufficient large to reach desired sensitivity as analyzed in Ref.~\cite{Chen:2023ryb}.  

\section{HFGW Constraint}
\begin{figure}[htbp]
\centering 
\includegraphics[width=1\textwidth]{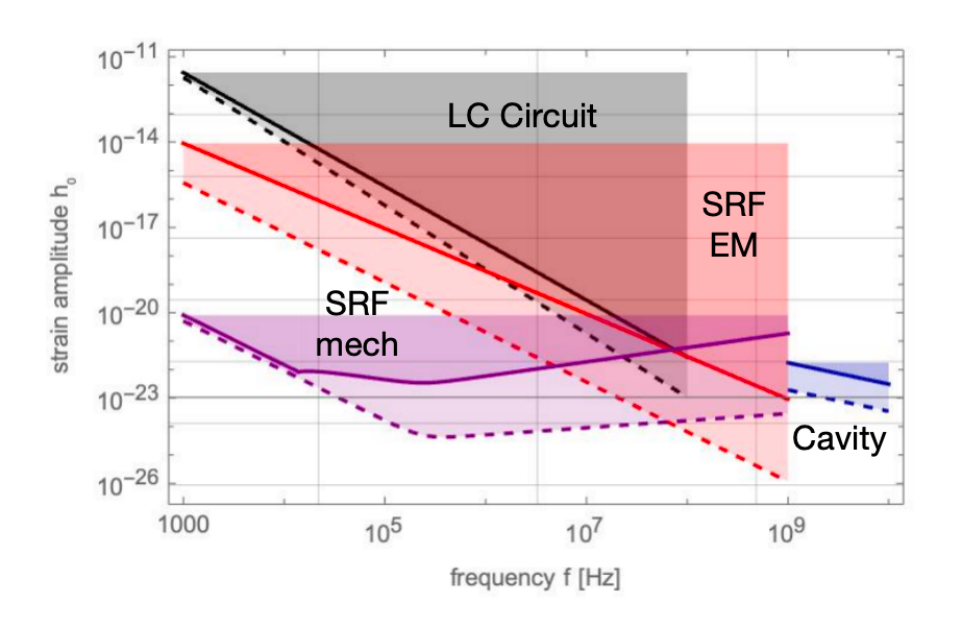}
\caption{\label{fig:HFGWconstraint}Sensitivity reach of HFGW for single-mode (solid) and multi-mode (dashed) detection limits.}
\end{figure}

By requiring $\mathrm{SNR}_{\rm single}^2=1$ and $\mathrm{SNR}_{\rm multiple}^2=1$ in Eqs.~\eqref{eq:single} and \eqref{eq:multiple}, respectively, we determine the strain sensitivity \(h_0(f=\omega_h/2\pi)\) for each detector, as shown in Fig.~\ref{fig:HFGWconstraint} (cf.\ Figure~2 of Ref.~\cite{Chen:2023ryb}). Figure~\ref{fig:HFGWconstraint} reveals:
\begin{enumerate}
  \item The multi‐mode (dashed) sensitivity exceeds the single‐mode (solid) sensitivity for all detectors.
  \item The LC circuit, SRF EM cavity and Cavity each reach their best sensitivity at their upper frequency bounds of \(10^8\)\,Hz, \(10^9\)\,Hz and \(10^{19}\)\,Hz, respectively, whereas the SRF mechanical resonator peaks around \(10^6\)\,Hz.
  \item Of all configurations, the SRF EM cavity achieves the highest strain sensitivity ($h_0=1.2\times 10^{-26}$ at $f=1~\mathrm{GHz}$).
\end{enumerate}
 
\begin{figure}[htbp]
\centering 
\includegraphics[width=1\textwidth]{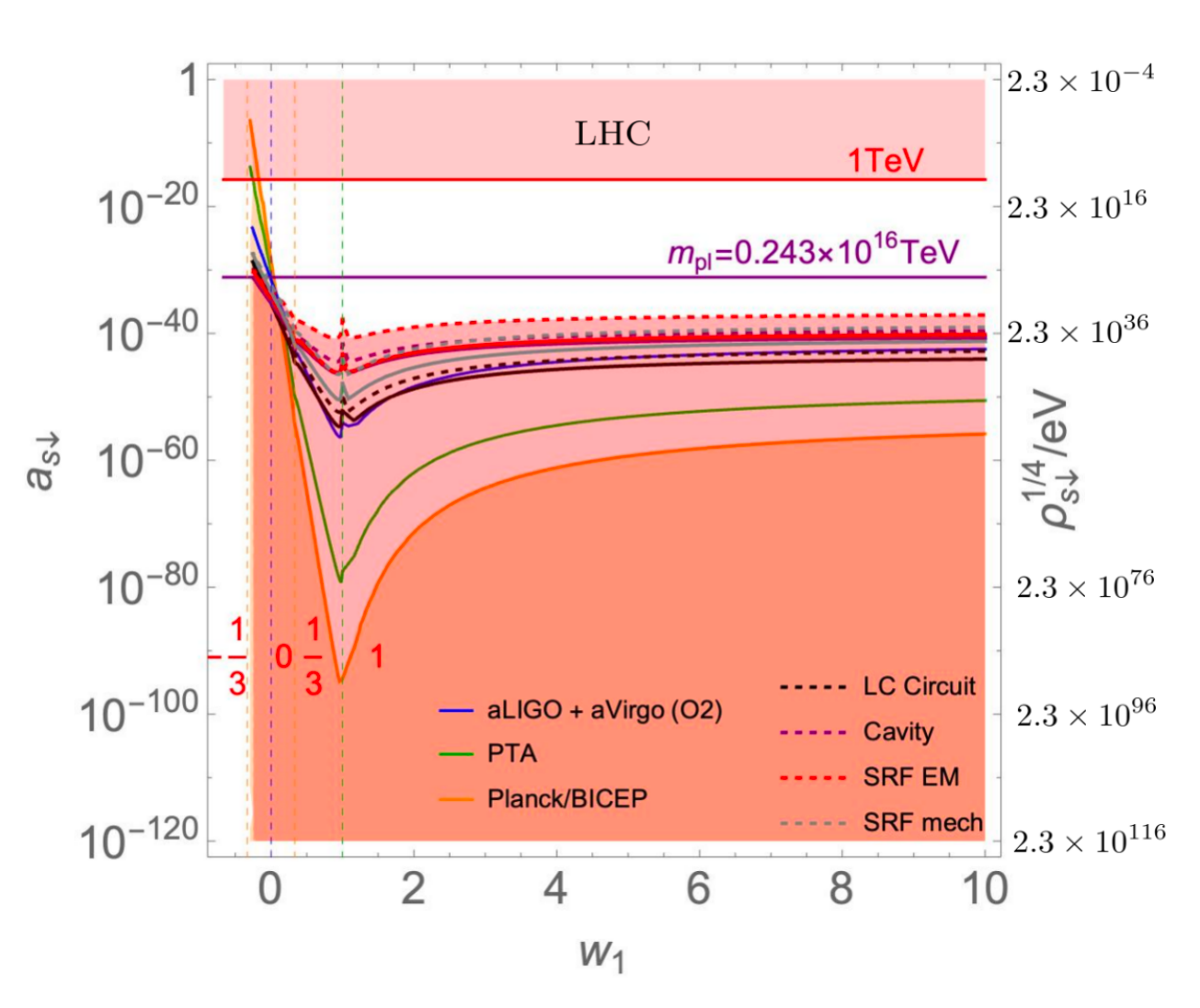}
\caption{\label{fig:LCSRF}HFGW constraints on the bounce energy scale of the generic bouncing universe for $w_1>-\frac{1}{3}$.}
\end{figure}

In Table~\ref{tab:sensitivity} we compile representative sensitivity values for each detector. Substituting these into Eq.~\eqref{eq:hzerorhos} yields the HFGW constraints on the bounce energy scale for a generic bouncing universe, as shown in Fig.~\ref{fig:LCSRF}. To generate Fig.~\ref{fig:LCSRF}, we employ the relation for the bounce scale factor,
\begin{equation}
a_{s\downarrow}\equiv a(\eta_{s\downarrow})
= \biggl(\frac{\rho_{c0}\,\Omega_{\gamma0}}{\rho_{s\downarrow}}\biggr)^{1/4},
\end{equation}
with $a_0=1$, and overlay three principal astrophysical GW bounds~\cite{Li:2025ilc}:  
$\Omega_{\rm GW}h^2=10^{-9}$ at $f=10^{-8}\,\mathrm{Hz}$ (PTAs: NANOGrav, PPTA, EPTA);  
$\Omega_{\rm GW}h^2=3.6\times10^{-17}$ (Planck/BICEP) at $f=7.75\times10^{-17}\,\mathrm{Hz}$;  
and $\Omega_{\rm GW}h^2=10^{-4}$ at $f=30\,\mathrm{Hz}$ (aLIGO–aVirgo O2). The lower limit $\rho_{s\downarrow}^{1/4}\gg1\ \mathrm{TeV}$ is motivated by the absence of any bounce‐triggering field at energies probed by the LHC.  For additional astrophysical GW constraints—from CMB‐S4, IPTA (design sensitivity), SKA, DECIGO, BBO, LISA, TianQin, Taiji, aLIGO+Virgo+KAGRA (design), Cosmic Explorer and the Einstein Telescope—see Refs.~\cite{Schmitz:2020syl,Annis:2022xgg,Bi:2023tib,Li:2025ilc}.

\begin{table}[htbp]
  \centering
  \caption{Detector sensitivities $h_0$ at representative frequencies}
  \label{tab:sensitivity}
  \begin{tabular}{l c c c}
    \toprule
    Detector & Frequency $f$ (Hz) 
      & {$h_{0,\mathrm{single}}$} 
      & {$h_{0,\mathrm{multi}}$} \\
    \midrule
    LC circuit 
      & $10^{7}$ 
      & $2.8\times10^{-20}$ 
      & $1.9\times10^{-21}$ \\
    Cavity 
      & $10^{10}$ 
      & $3.1\times10^{-23}$ 
      & $3.4\times10^{-24}$ \\
    SRF (EM mode) 
      & $10^{9}$ 
      & $9.1\times10^{-24}$ 
      & $1.2\times10^{-26}$ \\
    SRF (mechanical mode) 
      & $10^{6}$ 
      & $5.9\times10^{-23}$ 
      & $5.0\times10^{-25}$ \\
    \bottomrule
  \end{tabular}
\end{table}

In Fig.~\ref{fig:LCSRF}, we observe that for $-\tfrac{1}{3}<w_1\lesssim0$, the strongest bound on $\rho_{s\downarrow}^{1/4}$ arises from Planck/BICEP CMB measurements (orange solid), with the shaded regions below each curve excluded. Conversely, for $w_1>0$, HFGW detectors yield tighter constraints than all astrophysical probes. Notably, the SRF EM channel (red dashed) improves the limit on $\rho_{s\downarrow}^{1/4}$ by more than twenty orders of magnitude relative to Planck/BICEP in the regime $w_1\gg0$. This behavior reflects the spectral tilt of the SGWB: for $-\tfrac{1}{3}<w_1<0$, the spectrum is red‐tilted (see Eqs.~\eqref{eq:OGWrhos} and \eqref{eq:nt}), favoring low‐frequency experiments, whereas for $w_1>0$, it becomes blue‐tilted, enhancing the sensitivity of high‐frequency detectors.

\begin{figure}[htbp]
\centering 
\includegraphics[width=1\textwidth]{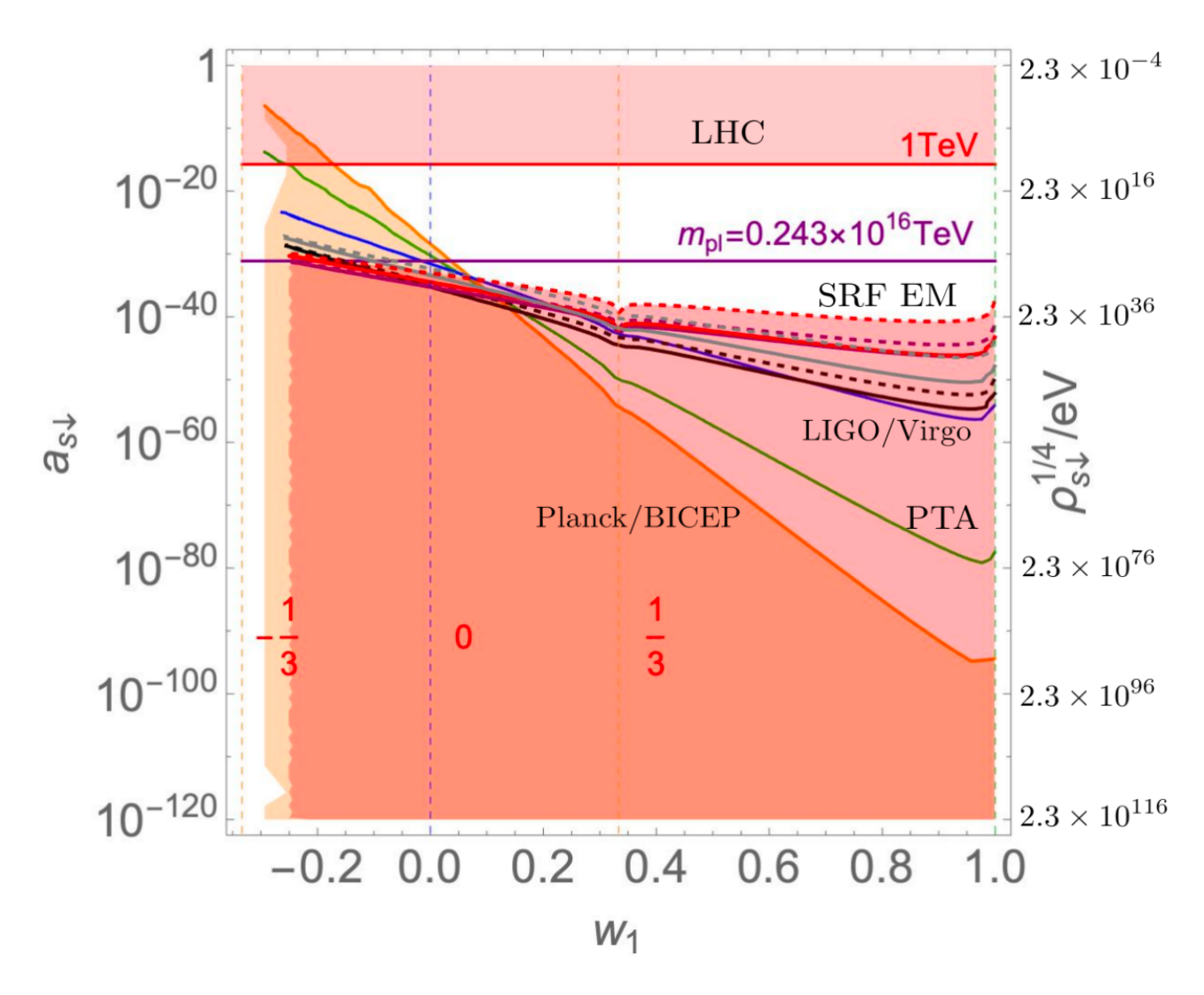}
\caption{\label{fig:LCSRFHighresolution}High resolution plot of Figure.~\ref{fig:LCSRF} for $-\frac{1}{3}<w_1\le 1$.}
\end{figure}

\begin{figure}[htbp]
\centering 
\includegraphics[width=1\textwidth]{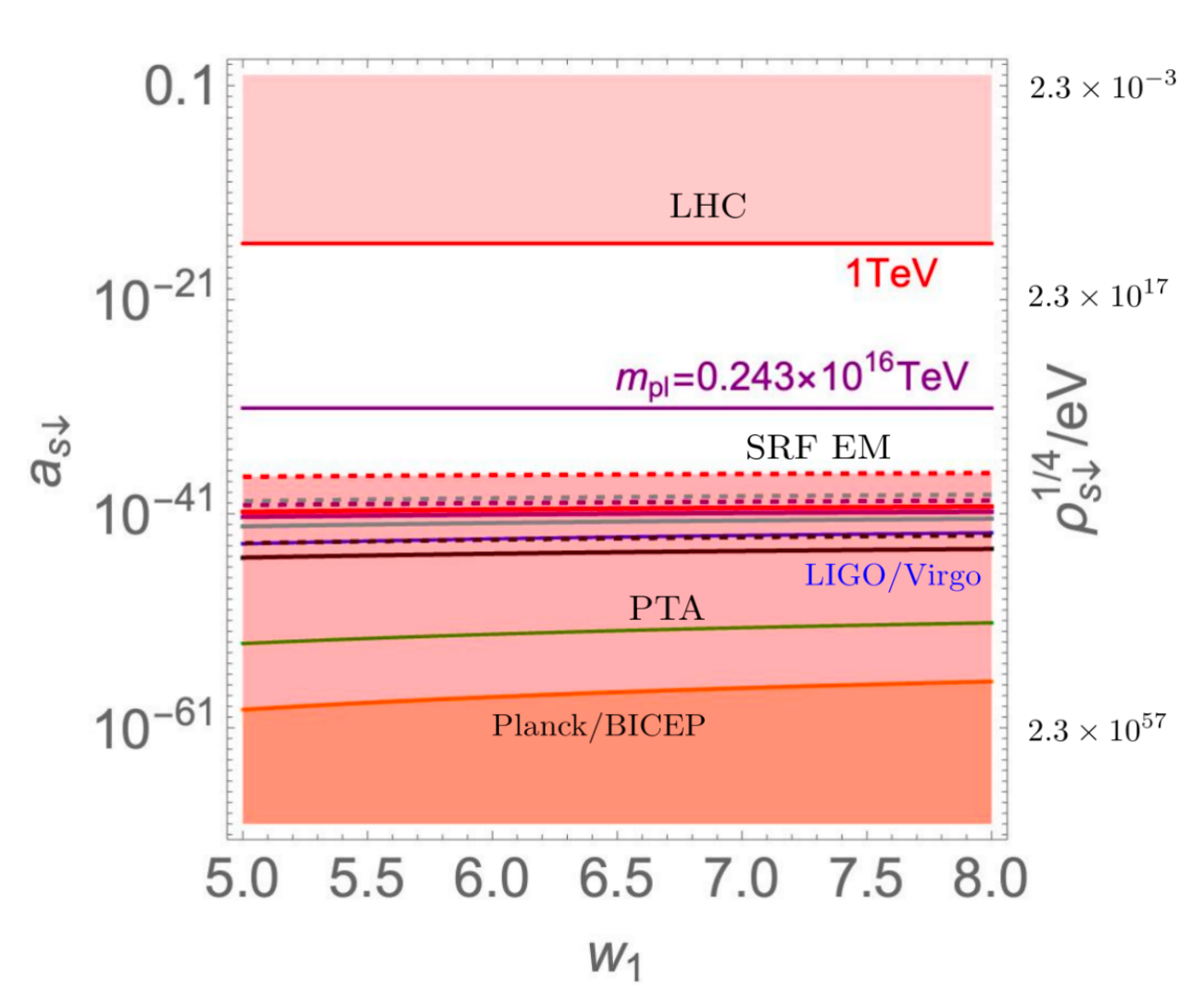}
\caption{\label{fig:LCSRFHighresolutionlarge}High resolution plot of Figure.~\ref{fig:LCSRF} for $5<w_1\le 8$.}
\end{figure}

In Figures~\ref{fig:LCSRFHighresolution} and \ref{fig:LCSRFHighresolutionlarge}, we present high‐resolution renditions of Fig.~\ref{fig:LCSRF}. Figure~\ref{fig:LCSRFHighresolution} demonstrates that all sensitivity curves converge at \(w_1\simeq0\), where the SGWB spectrum is scale‐invariant and both low‐ and high‐frequency detectors exhibit comparable reach. In cosmology, \(w_1=0\) corresponds to the matter‐bounce scenario, commonly realized via ghost‐condensation dynamics. For \(0\lesssim w_1<1\), the spectrum becomes blue‐tilted, so detectors with higher upper‐frequency coverage (e.g.\ aLIGO–aVirgo over PTAs, PTAs over CMB experiments) deliver progressively stronger constraints, with the SRF EM cavity achieving the best sensitivity. We further observe that for \(w_1>0.05\), neither current astrophysical probes nor resonant detectors can constrain \(\rho_{s\downarrow}^{1/4}\) down to the reduced Planck scale, and this feature persists in Fig.~\ref{fig:LCSRFHighresolutionlarge}. We prospect next‐generation LC circuits, traditional and SRF cavities with enhanced \(Q_{\mathrm{int}}\) and higher eigenfrequency \(\omega_{\mathrm{rf}}\) can impose tighter constrain in near future.

\section{Summary and Prospect}
In this work, we present the first constraints on the bounce energy scale \(\rho_{s\downarrow}^{1/4}\) within the generic bouncing‐cosmology framework over the frequency band \(1\,\mathrm{kHz}\lesssim f\lesssim10\,\mathrm{GHz}\), by exploiting forecast sensitivities to HFGWs from LC circuits, conventional resonant cavities, and SRF EM and mechanical cavities. We demonstrate that, relative to existing astrophysical GW limits from Planck/BICEP, PTAs and aLIGO+LISA+KAGRA, these HFGW detectors impose significantly tighter bounds on \(\rho_{s\downarrow}^{1/4}\) for \(w_1>0.05\). In particular, whereas Planck/BICEP provides the strongest constraint for \(-\tfrac{1}{3}<w_1<0\), the SRF EM channel improves the bound on \(\rho_{s\downarrow}^{1/4}\) by over twenty orders of magnitude in the regime \(w_1\gg0\). Our results highlight the power of HFGW resonators to probe key parameters of the very early Universe. By unifying astrophysical and quantum‐measurement constraints, we cover the GW spectrum from \(10^{-17}\,\mathrm{Hz}\) to \(10\,\mathrm{GHz}\), opening a new window on physics at unprecedented energy scales.

We further observe that, to further constrain the remaining parameter space of \(\rho_{s\downarrow}^{1/4}\), next‐generation LC circuits, conventional cavities and SRF (EM and mechanical) resonators must achieve intrinsic quality factors \(Q_{\rm int}\gg 10^{12}\) and operate at eigenfrequencies \(\omega_{\rm rf}\simeq w_0 \gg 10~\mathrm{GHz}\). Realizing such performance—particularly in superconducting devices, readout electronics, couplers and cryogenics at these frequencies—poses significant technological challenges, but we remain optimistic about future advances~\cite{Kahn:2023mrj,Chen:2023ryb,Domcke:2022rgu, Fang:2024ple,Berlin:2021txa, Franciolini:2022htd, Gong:2025xsd,Gatti:2024mde,SHANHE:2024tpr,Shu:2024nmc, Arakawa:2025hcn, Fischer:2024nte, Kohri:2024qpd, Yin:2025jqe, Zhong:2025pto,Li:2025eoo}.

\begin{acknowledgments}
C.L. is supported by the NSFC under Grants No.11963005 and No. 11603018, by Yunnan Provincial Foundation under Grants No.202401AT070459, No.2019FY003005, and No.2016FD006, by Young and Middle-aged Academic and Technical Leaders in Yunnan Province Program, by Yunnan Provincial High level Talent Training Support Plan Youth Top Program, by Yunnan University Donglu Talent Young Scholar, and by the NSFC under Grant No.11847301 and by the Fundamental Research Funds for the Central Universities under Grant No. 2019CDJDWL0005.
\end{acknowledgments}

\bibliography{biblio}

\begin{thebibliography}{102}
\expandafter\ifx\csname natexlab\endcsname\relax\def\natexlab#1{#1}\fi
\expandafter\ifx\csname bibnamefont\endcsname\relax
  \def\bibnamefont#1{#1}\fi
\expandafter\ifx\csname bibfnamefont\endcsname\relax
  \def\bibfnamefont#1{#1}\fi
\expandafter\ifx\csname citenamefont\endcsname\relax
  \def\citenamefont#1{#1}\fi
\expandafter\ifx\csname url\endcsname\relax
  \def\url#1{\texttt{#1}}\fi
\expandafter\ifx\csname urlprefix\endcsname\relax\def\urlprefix{URL }\fi
\providecommand{\bibinfo}[2]{#2}
\providecommand{\eprint}[2][]{\url{#2}}

\bibitem[{\citenamefont{Novello and Bergliaffa}(2008)}]{Novello:2008ra}
\bibinfo{author}{\bibfnamefont{M.}~\bibnamefont{Novello}} \bibnamefont{and} \bibinfo{author}{\bibfnamefont{S.~E.~P.} \bibnamefont{Bergliaffa}}, \bibinfo{journal}{Phys. Rept.} \textbf{\bibinfo{volume}{463}}, \bibinfo{pages}{127} (\bibinfo{year}{2008}), \eprint{0802.1634}.

\bibitem[{\citenamefont{Brandenberger and Peter}(2017)}]{Brandenberger:2016vhg}
\bibinfo{author}{\bibfnamefont{R.}~\bibnamefont{Brandenberger}} \bibnamefont{and} \bibinfo{author}{\bibfnamefont{P.}~\bibnamefont{Peter}}, \bibinfo{journal}{Found. Phys.} \textbf{\bibinfo{volume}{47}}, \bibinfo{pages}{797} (\bibinfo{year}{2017}), \eprint{1603.05834}.

\bibitem[{\citenamefont{Nojiri et~al.}(2017)\citenamefont{Nojiri, Odintsov, and Oikonomou}}]{Nojiri:2017ncd}
\bibinfo{author}{\bibfnamefont{S.}~\bibnamefont{Nojiri}}, \bibinfo{author}{\bibfnamefont{S.~D.} \bibnamefont{Odintsov}}, \bibnamefont{and} \bibinfo{author}{\bibfnamefont{V.~K.} \bibnamefont{Oikonomou}}, \bibinfo{journal}{Phys. Rept.} \textbf{\bibinfo{volume}{692}}, \bibinfo{pages}{1} (\bibinfo{year}{2017}), \eprint{1705.11098}.

\bibitem[{\citenamefont{Odintsov et~al.}(2023)\citenamefont{Odintsov, Oikonomou, Giannakoudi, Fronimos, and Lymperiadou}}]{Odintsov:2023weg}
\bibinfo{author}{\bibfnamefont{S.~D.} \bibnamefont{Odintsov}}, \bibinfo{author}{\bibfnamefont{V.~K.} \bibnamefont{Oikonomou}}, \bibinfo{author}{\bibfnamefont{I.}~\bibnamefont{Giannakoudi}}, \bibinfo{author}{\bibfnamefont{F.~P.} \bibnamefont{Fronimos}}, \bibnamefont{and} \bibinfo{author}{\bibfnamefont{E.~C.} \bibnamefont{Lymperiadou}}, \bibinfo{journal}{Symmetry} \textbf{\bibinfo{volume}{15}}, \bibinfo{pages}{1701} (\bibinfo{year}{2023}), \eprint{2307.16308}.

\bibitem[{\citenamefont{Khoury et~al.}(2001)\citenamefont{Khoury, Ovrut, Steinhardt, and Turok}}]{Khoury:2001wf}
\bibinfo{author}{\bibfnamefont{J.}~\bibnamefont{Khoury}}, \bibinfo{author}{\bibfnamefont{B.~A.} \bibnamefont{Ovrut}}, \bibinfo{author}{\bibfnamefont{P.~J.} \bibnamefont{Steinhardt}}, \bibnamefont{and} \bibinfo{author}{\bibfnamefont{N.}~\bibnamefont{Turok}}, \bibinfo{journal}{Phys. Rev. D} \textbf{\bibinfo{volume}{64}}, \bibinfo{pages}{123522} (\bibinfo{year}{2001}), \eprint{hep-th/0103239}.

\bibitem[{\citenamefont{Gasperini and Veneziano}(2003)}]{Gasperini:2002bn}
\bibinfo{author}{\bibfnamefont{M.}~\bibnamefont{Gasperini}} \bibnamefont{and} \bibinfo{author}{\bibfnamefont{G.}~\bibnamefont{Veneziano}}, \bibinfo{journal}{Phys. Rept.} \textbf{\bibinfo{volume}{373}}, \bibinfo{pages}{1} (\bibinfo{year}{2003}), \eprint{hep-th/0207130}.

\bibitem[{\citenamefont{Creminelli et~al.}(2006)\citenamefont{Creminelli, Luty, Nicolis, and Senatore}}]{Creminelli:2006xe}
\bibinfo{author}{\bibfnamefont{P.}~\bibnamefont{Creminelli}}, \bibinfo{author}{\bibfnamefont{M.~A.} \bibnamefont{Luty}}, \bibinfo{author}{\bibfnamefont{A.}~\bibnamefont{Nicolis}}, \bibnamefont{and} \bibinfo{author}{\bibfnamefont{L.}~\bibnamefont{Senatore}}, \bibinfo{journal}{JHEP} \textbf{\bibinfo{volume}{12}}, \bibinfo{pages}{080} (\bibinfo{year}{2006}), \eprint{hep-th/0606090}.

\bibitem[{\citenamefont{Peter et~al.}(2007)\citenamefont{Peter, Pinho, and Pinto-Neto}}]{Peter:2006hx}
\bibinfo{author}{\bibfnamefont{P.}~\bibnamefont{Peter}}, \bibinfo{author}{\bibfnamefont{E.~J.~C.} \bibnamefont{Pinho}}, \bibnamefont{and} \bibinfo{author}{\bibfnamefont{N.}~\bibnamefont{Pinto-Neto}}, \bibinfo{journal}{Phys. Rev. D} \textbf{\bibinfo{volume}{75}}, \bibinfo{pages}{023516} (\bibinfo{year}{2007}), \eprint{hep-th/0610205}.

\bibitem[{\citenamefont{Cai et~al.}(2007)\citenamefont{Cai, Qiu, Piao, Li, and Zhang}}]{Cai:2007qw}
\bibinfo{author}{\bibfnamefont{Y.-F.} \bibnamefont{Cai}}, \bibinfo{author}{\bibfnamefont{T.}~\bibnamefont{Qiu}}, \bibinfo{author}{\bibfnamefont{Y.-S.} \bibnamefont{Piao}}, \bibinfo{author}{\bibfnamefont{M.}~\bibnamefont{Li}}, \bibnamefont{and} \bibinfo{author}{\bibfnamefont{X.}~\bibnamefont{Zhang}}, \bibinfo{journal}{JHEP} \textbf{\bibinfo{volume}{10}}, \bibinfo{pages}{071} (\bibinfo{year}{2007}), \eprint{0704.1090}.

\bibitem[{\citenamefont{Cai et~al.}(2009{\natexlab{a}})\citenamefont{Cai, Qiu, Brandenberger, and Zhang}}]{Cai:2008qw}
\bibinfo{author}{\bibfnamefont{Y.-F.} \bibnamefont{Cai}}, \bibinfo{author}{\bibfnamefont{T.-t.} \bibnamefont{Qiu}}, \bibinfo{author}{\bibfnamefont{R.}~\bibnamefont{Brandenberger}}, \bibnamefont{and} \bibinfo{author}{\bibfnamefont{X.-m.} \bibnamefont{Zhang}}, \bibinfo{journal}{Phys. Rev. D} \textbf{\bibinfo{volume}{80}}, \bibinfo{pages}{023511} (\bibinfo{year}{2009}{\natexlab{a}}), \eprint{0810.4677}.

\bibitem[{\citenamefont{Saidov and Zhuk}(2010)}]{Saidov:2010wx}
\bibinfo{author}{\bibfnamefont{T.}~\bibnamefont{Saidov}} \bibnamefont{and} \bibinfo{author}{\bibfnamefont{A.}~\bibnamefont{Zhuk}}, \bibinfo{journal}{Phys. Rev. D} \textbf{\bibinfo{volume}{81}}, \bibinfo{pages}{124002} (\bibinfo{year}{2010}), \eprint{1002.4138}.

\bibitem[{\citenamefont{Li et~al.}(2014{\natexlab{a}})\citenamefont{Li, Wang, and Cheung}}]{Li:2011nj}
\bibinfo{author}{\bibfnamefont{C.}~\bibnamefont{Li}}, \bibinfo{author}{\bibfnamefont{L.}~\bibnamefont{Wang}}, \bibnamefont{and} \bibinfo{author}{\bibfnamefont{Y.-K.~E.} \bibnamefont{Cheung}}, \bibinfo{journal}{Phys. Dark Univ.} \textbf{\bibinfo{volume}{3}}, \bibinfo{pages}{18} (\bibinfo{year}{2014}{\natexlab{a}}), \eprint{1101.0202}.

\bibitem[{\citenamefont{Cai et~al.}(2011)\citenamefont{Cai, Chen, Dent, Dutta, and Saridakis}}]{Cai:2011tc}
\bibinfo{author}{\bibfnamefont{Y.-F.} \bibnamefont{Cai}}, \bibinfo{author}{\bibfnamefont{S.-H.} \bibnamefont{Chen}}, \bibinfo{author}{\bibfnamefont{J.~B.} \bibnamefont{Dent}}, \bibinfo{author}{\bibfnamefont{S.}~\bibnamefont{Dutta}}, \bibnamefont{and} \bibinfo{author}{\bibfnamefont{E.~N.} \bibnamefont{Saridakis}}, \bibinfo{journal}{Class. Quant. Grav.} \textbf{\bibinfo{volume}{28}}, \bibinfo{pages}{215011} (\bibinfo{year}{2011}), \eprint{1104.4349}.

\bibitem[{\citenamefont{Easson et~al.}(2011)\citenamefont{Easson, Sawicki, and Vikman}}]{Easson:2011zy}
\bibinfo{author}{\bibfnamefont{D.~A.} \bibnamefont{Easson}}, \bibinfo{author}{\bibfnamefont{I.}~\bibnamefont{Sawicki}}, \bibnamefont{and} \bibinfo{author}{\bibfnamefont{A.}~\bibnamefont{Vikman}}, \bibinfo{journal}{JCAP} \textbf{\bibinfo{volume}{11}}, \bibinfo{pages}{021} (\bibinfo{year}{2011}), \eprint{1109.1047}.

\bibitem[{\citenamefont{Bhattacharya et~al.}(2013)\citenamefont{Bhattacharya, Cai, and Das}}]{Bhattacharya:2013ut}
\bibinfo{author}{\bibfnamefont{K.}~\bibnamefont{Bhattacharya}}, \bibinfo{author}{\bibfnamefont{Y.-F.} \bibnamefont{Cai}}, \bibnamefont{and} \bibinfo{author}{\bibfnamefont{S.}~\bibnamefont{Das}}, \bibinfo{journal}{Phys. Rev. D} \textbf{\bibinfo{volume}{87}}, \bibinfo{pages}{083511} (\bibinfo{year}{2013}), \eprint{1301.0661}.

\bibitem[{\citenamefont{Qiu and Wang}(2015)}]{Qiu:2015nha}
\bibinfo{author}{\bibfnamefont{T.}~\bibnamefont{Qiu}} \bibnamefont{and} \bibinfo{author}{\bibfnamefont{Y.-T.} \bibnamefont{Wang}}, \bibinfo{journal}{JHEP} \textbf{\bibinfo{volume}{04}}, \bibinfo{pages}{130} (\bibinfo{year}{2015}), \eprint{1501.03568}.

\bibitem[{\citenamefont{Barrow and Ganguly}(2017)}]{Barrow:2017yqt}
\bibinfo{author}{\bibfnamefont{J.~D.} \bibnamefont{Barrow}} \bibnamefont{and} \bibinfo{author}{\bibfnamefont{C.}~\bibnamefont{Ganguly}}, \bibinfo{journal}{Phys. Rev. D} \textbf{\bibinfo{volume}{95}}, \bibinfo{pages}{083515} (\bibinfo{year}{2017}), \eprint{1703.05969}.

\bibitem[{\citenamefont{de~Haro and Amor\'os}(2018)}]{deHaro:2017yll}
\bibinfo{author}{\bibfnamefont{J.}~\bibnamefont{de~Haro}} \bibnamefont{and} \bibinfo{author}{\bibfnamefont{J.}~\bibnamefont{Amor\'os}}, \bibinfo{journal}{Phys. Rev. D} \textbf{\bibinfo{volume}{97}}, \bibinfo{pages}{064014} (\bibinfo{year}{2018}), \eprint{1712.08399}.

\bibitem[{\citenamefont{Boruah et~al.}(2018)\citenamefont{Boruah, Kim, Rouben, and Geshnizjani}}]{Boruah:2018pvq}
\bibinfo{author}{\bibfnamefont{S.~S.} \bibnamefont{Boruah}}, \bibinfo{author}{\bibfnamefont{H.~J.} \bibnamefont{Kim}}, \bibinfo{author}{\bibfnamefont{M.}~\bibnamefont{Rouben}}, \bibnamefont{and} \bibinfo{author}{\bibfnamefont{G.}~\bibnamefont{Geshnizjani}}, \bibinfo{journal}{JCAP} \textbf{\bibinfo{volume}{08}}, \bibinfo{pages}{031} (\bibinfo{year}{2018}), \eprint{1802.06818}.

\bibitem[{\citenamefont{Nojiri et~al.}(2019)\citenamefont{Nojiri, Odintsov, and Saridakis}}]{Nojiri:2019yzg}
\bibinfo{author}{\bibfnamefont{S.}~\bibnamefont{Nojiri}}, \bibinfo{author}{\bibfnamefont{S.~D.} \bibnamefont{Odintsov}}, \bibnamefont{and} \bibinfo{author}{\bibfnamefont{E.~N.} \bibnamefont{Saridakis}}, \bibinfo{journal}{Nucl. Phys. B} \textbf{\bibinfo{volume}{949}}, \bibinfo{pages}{114790} (\bibinfo{year}{2019}), \eprint{1908.00389}.

\bibitem[{\citenamefont{Silva}(2018)}]{Silva:2015qna}
\bibinfo{author}{\bibfnamefont{C.~A.~S.} \bibnamefont{Silva}}, \bibinfo{journal}{Eur. Phys. J. C} \textbf{\bibinfo{volume}{78}}, \bibinfo{pages}{409} (\bibinfo{year}{2018}), \eprint{1503.00559}.

\bibitem[{\citenamefont{Silva}(2020)}]{Silva:2020bnn}
\bibinfo{author}{\bibfnamefont{C.}~\bibnamefont{Silva}}, \bibinfo{journal}{Phys. Rev. D} \textbf{\bibinfo{volume}{102}}, \bibinfo{pages}{046001} (\bibinfo{year}{2020}), \eprint{2008.07279}.

\bibitem[{\citenamefont{Silva}(2024)}]{Silva:2023ieb}
\bibinfo{author}{\bibfnamefont{C.}~\bibnamefont{Silva}}, \bibinfo{journal}{Nucl. Phys. B} \textbf{\bibinfo{volume}{998}}, \bibinfo{pages}{116402} (\bibinfo{year}{2024}), \eprint{2312.05260}.

\bibitem[{\citenamefont{Tukhashvili and Steinhardt}(2023)}]{Tukhashvili:2023itb}
\bibinfo{author}{\bibfnamefont{G.}~\bibnamefont{Tukhashvili}} \bibnamefont{and} \bibinfo{author}{\bibfnamefont{P.~J.} \bibnamefont{Steinhardt}}, \bibinfo{journal}{Phys. Rev. Lett.} \textbf{\bibinfo{volume}{131}}, \bibinfo{pages}{091001} (\bibinfo{year}{2023}), \eprint{2307.16098}.

\bibitem[{\citenamefont{Garcia-Saenz et~al.}(2024)\citenamefont{Garcia-Saenz, Hua, and Zhao}}]{Garcia-Saenz:2024ogr}
\bibinfo{author}{\bibfnamefont{S.}~\bibnamefont{Garcia-Saenz}}, \bibinfo{author}{\bibfnamefont{J.}~\bibnamefont{Hua}}, \bibnamefont{and} \bibinfo{author}{\bibfnamefont{Y.}~\bibnamefont{Zhao}}, \bibinfo{journal}{Phys. Rev. D} \textbf{\bibinfo{volume}{110}}, \bibinfo{pages}{L061304} (\bibinfo{year}{2024}), \eprint{2405.04062}.

\bibitem[{\citenamefont{Li et~al.}(2025{\natexlab{a}})\citenamefont{Li, Zhang, and Qiu}}]{Li:2024rgq}
\bibinfo{author}{\bibfnamefont{H.-H.} \bibnamefont{Li}}, \bibinfo{author}{\bibfnamefont{X.-z.} \bibnamefont{Zhang}}, \bibnamefont{and} \bibinfo{author}{\bibfnamefont{T.}~\bibnamefont{Qiu}}, \bibinfo{journal}{Sci. Bull.} \textbf{\bibinfo{volume}{70}}, \bibinfo{pages}{829} (\bibinfo{year}{2025}{\natexlab{a}}), \eprint{2409.04027}.

\bibitem[{\citenamefont{Nayeri et~al.}(2006)\citenamefont{Nayeri, Brandenberger, and Vafa}}]{Nayeri:2005ck}
\bibinfo{author}{\bibfnamefont{A.}~\bibnamefont{Nayeri}}, \bibinfo{author}{\bibfnamefont{R.~H.} \bibnamefont{Brandenberger}}, \bibnamefont{and} \bibinfo{author}{\bibfnamefont{C.}~\bibnamefont{Vafa}}, \bibinfo{journal}{Phys. Rev. Lett.} \textbf{\bibinfo{volume}{97}}, \bibinfo{pages}{021302} (\bibinfo{year}{2006}), \eprint{hep-th/0511140}.

\bibitem[{\citenamefont{Brandenberger et~al.}(2007)\citenamefont{Brandenberger, Nayeri, Patil, and Vafa}}]{Brandenberger:2006xi}
\bibinfo{author}{\bibfnamefont{R.~H.} \bibnamefont{Brandenberger}}, \bibinfo{author}{\bibfnamefont{A.}~\bibnamefont{Nayeri}}, \bibinfo{author}{\bibfnamefont{S.~P.} \bibnamefont{Patil}}, \bibnamefont{and} \bibinfo{author}{\bibfnamefont{C.}~\bibnamefont{Vafa}}, \bibinfo{journal}{Phys. Rev. Lett.} \textbf{\bibinfo{volume}{98}}, \bibinfo{pages}{231302} (\bibinfo{year}{2007}), \eprint{hep-th/0604126}.

\bibitem[{\citenamefont{Fischler and Susskind}(1998)}]{Fischler:1998st}
\bibinfo{author}{\bibfnamefont{W.}~\bibnamefont{Fischler}} \bibnamefont{and} \bibinfo{author}{\bibfnamefont{L.}~\bibnamefont{Susskind}} (\bibinfo{year}{1998}), \eprint{hep-th/9806039}.

\bibitem[{\citenamefont{Cai et~al.}(2009{\natexlab{b}})\citenamefont{Cai, Xue, Brandenberger, and Zhang}}]{Cai:2009rd}
\bibinfo{author}{\bibfnamefont{Y.-F.} \bibnamefont{Cai}}, \bibinfo{author}{\bibfnamefont{W.}~\bibnamefont{Xue}}, \bibinfo{author}{\bibfnamefont{R.}~\bibnamefont{Brandenberger}}, \bibnamefont{and} \bibinfo{author}{\bibfnamefont{X.-m.} \bibnamefont{Zhang}}, \bibinfo{journal}{JCAP} \textbf{\bibinfo{volume}{06}}, \bibinfo{pages}{037} (\bibinfo{year}{2009}{\natexlab{b}}), \eprint{0903.4938}.

\bibitem[{\citenamefont{Cheung et~al.}(2014)\citenamefont{Cheung, Kang, and Li}}]{Cheung:2014nxi}
\bibinfo{author}{\bibfnamefont{Y.-K.~E.} \bibnamefont{Cheung}}, \bibinfo{author}{\bibfnamefont{J.~U.} \bibnamefont{Kang}}, \bibnamefont{and} \bibinfo{author}{\bibfnamefont{C.}~\bibnamefont{Li}}, \bibinfo{journal}{JCAP} \textbf{\bibinfo{volume}{11}}, \bibinfo{pages}{001} (\bibinfo{year}{2014}), \eprint{1408.4387}.

\bibitem[{\citenamefont{Li}(2015)}]{Li:2014cba}
\bibinfo{author}{\bibfnamefont{C.}~\bibnamefont{Li}}, \bibinfo{journal}{Phys. Rev. D} \textbf{\bibinfo{volume}{92}}, \bibinfo{pages}{063513} (\bibinfo{year}{2015}), \eprint{1404.4012}.

\bibitem[{\citenamefont{Li}(2016)}]{Li:2015egy}
\bibinfo{author}{\bibfnamefont{C.}~\bibnamefont{Li}}, \bibinfo{journal}{JCAP} \textbf{\bibinfo{volume}{09}}, \bibinfo{pages}{038} (\bibinfo{year}{2016}), \eprint{1512.06794}.

\bibitem[{\citenamefont{Guth}(1981)}]{Guth:1980zm}
\bibinfo{author}{\bibfnamefont{A.~H.} \bibnamefont{Guth}}, \bibinfo{journal}{Phys. Rev. D} \textbf{\bibinfo{volume}{23}}, \bibinfo{pages}{347} (\bibinfo{year}{1981}).

\bibitem[{\citenamefont{Starobinsky}(1980)}]{Starobinsky:1980te}
\bibinfo{author}{\bibfnamefont{A.~A.} \bibnamefont{Starobinsky}}, \bibinfo{journal}{Phys. Lett. B} \textbf{\bibinfo{volume}{91}}, \bibinfo{pages}{99} (\bibinfo{year}{1980}).

\bibitem[{\citenamefont{Sato}(1981)}]{Sato:1980yn}
\bibinfo{author}{\bibfnamefont{K.}~\bibnamefont{Sato}}, \bibinfo{journal}{Mon. Not. Roy. Astron. Soc.} \textbf{\bibinfo{volume}{195}}, \bibinfo{pages}{467} (\bibinfo{year}{1981}).

\bibitem[{\citenamefont{Linde}(1982)}]{Linde:1981mu}
\bibinfo{author}{\bibfnamefont{A.~D.} \bibnamefont{Linde}}, \bibinfo{journal}{Phys. Lett. B} \textbf{\bibinfo{volume}{108}}, \bibinfo{pages}{389} (\bibinfo{year}{1982}).

\bibitem[{\citenamefont{Albrecht and Steinhardt}(1982)}]{Albrecht:1982wi}
\bibinfo{author}{\bibfnamefont{A.}~\bibnamefont{Albrecht}} \bibnamefont{and} \bibinfo{author}{\bibfnamefont{P.~J.} \bibnamefont{Steinhardt}}, \bibinfo{journal}{Phys. Rev. Lett.} \textbf{\bibinfo{volume}{48}}, \bibinfo{pages}{1220} (\bibinfo{year}{1982}).

\bibitem[{\citenamefont{Mukhanov et~al.}(1992)\citenamefont{Mukhanov, Feldman, and Brandenberger}}]{Mukhanov:1990me}
\bibinfo{author}{\bibfnamefont{V.~F.} \bibnamefont{Mukhanov}}, \bibinfo{author}{\bibfnamefont{H.~A.} \bibnamefont{Feldman}}, \bibnamefont{and} \bibinfo{author}{\bibfnamefont{R.~H.} \bibnamefont{Brandenberger}}, \bibinfo{journal}{Phys. Rept.} \textbf{\bibinfo{volume}{215}}, \bibinfo{pages}{203} (\bibinfo{year}{1992}).

\bibitem[{\citenamefont{Borde and Vilenkin}(1994)}]{Borde:1993xh}
\bibinfo{author}{\bibfnamefont{A.}~\bibnamefont{Borde}} \bibnamefont{and} \bibinfo{author}{\bibfnamefont{A.}~\bibnamefont{Vilenkin}}, \bibinfo{journal}{Phys. Rev. Lett.} \textbf{\bibinfo{volume}{72}}, \bibinfo{pages}{3305} (\bibinfo{year}{1994}), \eprint{gr-qc/9312022}.

\bibitem[{\citenamefont{Borde et~al.}(2003)\citenamefont{Borde, Guth, and Vilenkin}}]{Borde:2001nh}
\bibinfo{author}{\bibfnamefont{A.}~\bibnamefont{Borde}}, \bibinfo{author}{\bibfnamefont{A.~H.} \bibnamefont{Guth}}, \bibnamefont{and} \bibinfo{author}{\bibfnamefont{A.}~\bibnamefont{Vilenkin}}, \bibinfo{journal}{Phys. Rev. Lett.} \textbf{\bibinfo{volume}{90}}, \bibinfo{pages}{151301} (\bibinfo{year}{2003}), \eprint{gr-qc/0110012}.

\bibitem[{\citenamefont{Ijjas and Steinhardt}(2018)}]{Ijjas:2018qbo}
\bibinfo{author}{\bibfnamefont{A.}~\bibnamefont{Ijjas}} \bibnamefont{and} \bibinfo{author}{\bibfnamefont{P.~J.} \bibnamefont{Steinhardt}}, \bibinfo{journal}{Class. Quant. Grav.} \textbf{\bibinfo{volume}{35}}, \bibinfo{pages}{135004} (\bibinfo{year}{2018}), \eprint{1803.01961}.

\bibitem[{\citenamefont{Komatsu et~al.}(2011)}]{WMAP:2010qai}
\bibinfo{author}{\bibfnamefont{E.}~\bibnamefont{Komatsu}} \bibnamefont{et~al.} (\bibinfo{collaboration}{WMAP}), \bibinfo{journal}{Astrophys. J. Suppl.} \textbf{\bibinfo{volume}{192}}, \bibinfo{pages}{18} (\bibinfo{year}{2011}), \eprint{1001.4538}.

\bibitem[{\citenamefont{Ade et~al.}(2016)}]{Planck:2015fie}
\bibinfo{author}{\bibfnamefont{P.~A.~R.} \bibnamefont{Ade}} \bibnamefont{et~al.} (\bibinfo{collaboration}{Planck}), \bibinfo{journal}{Astron. Astrophys.} \textbf{\bibinfo{volume}{594}}, \bibinfo{pages}{A13} (\bibinfo{year}{2016}), \eprint{1502.01589}.

\bibitem[{\citenamefont{Aghanim et~al.}(2020)}]{Planck:2018vyg}
\bibinfo{author}{\bibfnamefont{N.}~\bibnamefont{Aghanim}} \bibnamefont{et~al.} (\bibinfo{collaboration}{Planck}), \bibinfo{journal}{Astron. Astrophys.} \textbf{\bibinfo{volume}{641}}, \bibinfo{pages}{A6} (\bibinfo{year}{2020}), \bibinfo{note}{[Erratum: Astron.Astrophys. 652, C4 (2021)]}, \eprint{1807.06209}.

\bibitem[{\citenamefont{Li and Cheung}(2014)}]{Li:2013bha}
\bibinfo{author}{\bibfnamefont{C.}~\bibnamefont{Li}} \bibnamefont{and} \bibinfo{author}{\bibfnamefont{Y.-K.~E.} \bibnamefont{Cheung}}, \bibinfo{journal}{JCAP} \textbf{\bibinfo{volume}{07}}, \bibinfo{pages}{008} (\bibinfo{year}{2014}), \eprint{1401.0094}.

\bibitem[{\citenamefont{Barrie}(2021)}]{Barrie:2021orn}
\bibinfo{author}{\bibfnamefont{N.~D.} \bibnamefont{Barrie}}, \bibinfo{journal}{JCAP} \textbf{\bibinfo{volume}{06}}, \bibinfo{pages}{049} (\bibinfo{year}{2021}), \eprint{2105.06624}.

\bibitem[{\citenamefont{Li et~al.}(2014{\natexlab{b}})\citenamefont{Li, Brandenberger, and Cheung}}]{Li:2014era}
\bibinfo{author}{\bibfnamefont{C.}~\bibnamefont{Li}}, \bibinfo{author}{\bibfnamefont{R.~H.} \bibnamefont{Brandenberger}}, \bibnamefont{and} \bibinfo{author}{\bibfnamefont{Y.-K.~E.} \bibnamefont{Cheung}}, \bibinfo{journal}{Phys. Rev. D} \textbf{\bibinfo{volume}{90}}, \bibinfo{pages}{123535} (\bibinfo{year}{2014}{\natexlab{b}}), \eprint{1403.5625}.

\bibitem[{\citenamefont{Li}(2020)}]{Li:2020nah}
\bibinfo{author}{\bibfnamefont{C.}~\bibnamefont{Li}}, \bibinfo{journal}{Phys. Rev. D} \textbf{\bibinfo{volume}{102}}, \bibinfo{pages}{123530} (\bibinfo{year}{2020}), \eprint{2008.10264}.

\bibitem[{\citenamefont{Agazie et~al.}(2023)}]{NANOGrav:2023gor}
\bibinfo{author}{\bibfnamefont{G.}~\bibnamefont{Agazie}} \bibnamefont{et~al.} (\bibinfo{collaboration}{NANOGrav}), \bibinfo{journal}{Astrophys. J. Lett.} \textbf{\bibinfo{volume}{951}}, \bibinfo{pages}{L8} (\bibinfo{year}{2023}), \eprint{2306.16213}.

\bibitem[{\citenamefont{Antoniadis et~al.}(2023)}]{EPTA:2023fyk}
\bibinfo{author}{\bibfnamefont{J.}~\bibnamefont{Antoniadis}} \bibnamefont{et~al.} (\bibinfo{collaboration}{EPTA, InPTA:}), \bibinfo{journal}{Astron. Astrophys.} \textbf{\bibinfo{volume}{678}}, \bibinfo{pages}{A50} (\bibinfo{year}{2023}), \eprint{2306.16214}.

\bibitem[{\citenamefont{Reardon et~al.}(2023)}]{Reardon:2023gzh}
\bibinfo{author}{\bibfnamefont{D.~J.} \bibnamefont{Reardon}} \bibnamefont{et~al.}, \bibinfo{journal}{Astrophys. J. Lett.} \textbf{\bibinfo{volume}{951}}, \bibinfo{pages}{L6} (\bibinfo{year}{2023}), \eprint{2306.16215}.

\bibitem[{\citenamefont{Antoniadis et~al.}(2022)}]{Antoniadis:2022pcn}
\bibinfo{author}{\bibfnamefont{J.}~\bibnamefont{Antoniadis}} \bibnamefont{et~al.}, \bibinfo{journal}{Mon. Not. Roy. Astron. Soc.} \textbf{\bibinfo{volume}{510}}, \bibinfo{pages}{4873} (\bibinfo{year}{2022}), \eprint{2201.03980}.

\bibitem[{\citenamefont{Xu et~al.}(2023)}]{Xu:2023wog}
\bibinfo{author}{\bibfnamefont{H.}~\bibnamefont{Xu}} \bibnamefont{et~al.}, \bibinfo{journal}{Res. Astron. Astrophys.} \textbf{\bibinfo{volume}{23}}, \bibinfo{pages}{075024} (\bibinfo{year}{2023}), \eprint{2306.16216}.

\bibitem[{\citenamefont{Afzal et~al.}(2023)}]{NANOGrav:2023hvm}
\bibinfo{author}{\bibfnamefont{A.}~\bibnamefont{Afzal}} \bibnamefont{et~al.} (\bibinfo{collaboration}{NANOGrav}), \bibinfo{journal}{Astrophys. J. Lett.} \textbf{\bibinfo{volume}{951}}, \bibinfo{pages}{L11} (\bibinfo{year}{2023}), \eprint{2306.16219}.

\bibitem[{\citenamefont{Antoniadis et~al.}(2024)}]{EPTA:2023xxk}
\bibinfo{author}{\bibfnamefont{J.}~\bibnamefont{Antoniadis}} \bibnamefont{et~al.} (\bibinfo{collaboration}{EPTA, InPTA}), \bibinfo{journal}{Astron. Astrophys.} \textbf{\bibinfo{volume}{685}}, \bibinfo{pages}{A94} (\bibinfo{year}{2024}), \eprint{2306.16227}.

\bibitem[{\citenamefont{Figueroa et~al.}(2024)\citenamefont{Figueroa, Pieroni, Ricciardone, and Simakachorn}}]{Figueroa:2023zhu}
\bibinfo{author}{\bibfnamefont{D.~G.} \bibnamefont{Figueroa}}, \bibinfo{author}{\bibfnamefont{M.}~\bibnamefont{Pieroni}}, \bibinfo{author}{\bibfnamefont{A.}~\bibnamefont{Ricciardone}}, \bibnamefont{and} \bibinfo{author}{\bibfnamefont{P.}~\bibnamefont{Simakachorn}}, \bibinfo{journal}{Phys. Rev. Lett.} \textbf{\bibinfo{volume}{132}}, \bibinfo{pages}{171002} (\bibinfo{year}{2024}), \eprint{2307.02399}.

\bibitem[{\citenamefont{Bian et~al.}(2024)\citenamefont{Bian, Ge, Shu, Wang, Yang, and Zong}}]{Bian:2023dnv}
\bibinfo{author}{\bibfnamefont{L.}~\bibnamefont{Bian}}, \bibinfo{author}{\bibfnamefont{S.}~\bibnamefont{Ge}}, \bibinfo{author}{\bibfnamefont{J.}~\bibnamefont{Shu}}, \bibinfo{author}{\bibfnamefont{B.}~\bibnamefont{Wang}}, \bibinfo{author}{\bibfnamefont{X.-Y.} \bibnamefont{Yang}}, \bibnamefont{and} \bibinfo{author}{\bibfnamefont{J.}~\bibnamefont{Zong}}, \bibinfo{journal}{Phys. Rev. D} \textbf{\bibinfo{volume}{109}}, \bibinfo{pages}{L101301} (\bibinfo{year}{2024}), \eprint{2307.02376}.

\bibitem[{\citenamefont{Ellis et~al.}(2024)\citenamefont{Ellis, Fairbairn, Franciolini, H\"utsi, Iovino, Lewicki, Raidal, Urrutia, Vaskonen, and Veerm\"ae}}]{Ellis:2023oxs}
\bibinfo{author}{\bibfnamefont{J.}~\bibnamefont{Ellis}}, \bibinfo{author}{\bibfnamefont{M.}~\bibnamefont{Fairbairn}}, \bibinfo{author}{\bibfnamefont{G.}~\bibnamefont{Franciolini}}, \bibinfo{author}{\bibfnamefont{G.}~\bibnamefont{H\"utsi}}, \bibinfo{author}{\bibfnamefont{A.}~\bibnamefont{Iovino}}, \bibinfo{author}{\bibfnamefont{M.}~\bibnamefont{Lewicki}}, \bibinfo{author}{\bibfnamefont{M.}~\bibnamefont{Raidal}}, \bibinfo{author}{\bibfnamefont{J.}~\bibnamefont{Urrutia}}, \bibinfo{author}{\bibfnamefont{V.}~\bibnamefont{Vaskonen}}, \bibnamefont{and} \bibinfo{author}{\bibfnamefont{H.}~\bibnamefont{Veerm\"ae}}, \bibinfo{journal}{Phys. Rev. D} \textbf{\bibinfo{volume}{109}}, \bibinfo{pages}{023522} (\bibinfo{year}{2024}), \eprint{2308.08546}.

\bibitem[{\citenamefont{Lai and Li}(2025{\natexlab{a}})}]{Lai:2025xov}
\bibinfo{author}{\bibfnamefont{J.}~\bibnamefont{Lai}} \bibnamefont{and} \bibinfo{author}{\bibfnamefont{C.}~\bibnamefont{Li}} (\bibinfo{year}{2025}{\natexlab{a}}), \eprint{2504.04211}.

\bibitem[{\citenamefont{Caprini and Figueroa}(2018)}]{Caprini:2018mtu}
\bibinfo{author}{\bibfnamefont{C.}~\bibnamefont{Caprini}} \bibnamefont{and} \bibinfo{author}{\bibfnamefont{D.~G.} \bibnamefont{Figueroa}}, \bibinfo{journal}{Class. Quant. Grav.} \textbf{\bibinfo{volume}{35}}, \bibinfo{pages}{163001} (\bibinfo{year}{2018}), \eprint{1801.04268}.

\bibitem[{\citenamefont{Zhao et~al.}(2013)\citenamefont{Zhao, Zhang, You, and Zhu}}]{Zhao:2013bba}
\bibinfo{author}{\bibfnamefont{W.}~\bibnamefont{Zhao}}, \bibinfo{author}{\bibfnamefont{Y.}~\bibnamefont{Zhang}}, \bibinfo{author}{\bibfnamefont{X.-P.} \bibnamefont{You}}, \bibnamefont{and} \bibinfo{author}{\bibfnamefont{Z.-H.} \bibnamefont{Zhu}}, \bibinfo{journal}{Phys. Rev. D} \textbf{\bibinfo{volume}{87}}, \bibinfo{pages}{124012} (\bibinfo{year}{2013}), \eprint{1303.6718}.

\bibitem[{\citenamefont{Guzzetti et~al.}(2016)\citenamefont{Guzzetti, Bartolo, Liguori, and Matarrese}}]{Guzzetti:2016mkm}
\bibinfo{author}{\bibfnamefont{M.~C.} \bibnamefont{Guzzetti}}, \bibinfo{author}{\bibfnamefont{N.}~\bibnamefont{Bartolo}}, \bibinfo{author}{\bibfnamefont{M.}~\bibnamefont{Liguori}}, \bibnamefont{and} \bibinfo{author}{\bibfnamefont{S.}~\bibnamefont{Matarrese}}, \bibinfo{journal}{Riv. Nuovo Cim.} \textbf{\bibinfo{volume}{39}}, \bibinfo{pages}{399} (\bibinfo{year}{2016}), \eprint{1605.01615}.

\bibitem[{\citenamefont{Cai et~al.}(2016)\citenamefont{Cai, Marciano, Wang, and Wilson-Ewing}}]{Cai:2016hea}
\bibinfo{author}{\bibfnamefont{Y.-F.} \bibnamefont{Cai}}, \bibinfo{author}{\bibfnamefont{A.}~\bibnamefont{Marciano}}, \bibinfo{author}{\bibfnamefont{D.-G.} \bibnamefont{Wang}}, \bibnamefont{and} \bibinfo{author}{\bibfnamefont{E.}~\bibnamefont{Wilson-Ewing}}, \bibinfo{journal}{Universe} \textbf{\bibinfo{volume}{3}}, \bibinfo{pages}{1} (\bibinfo{year}{2016}), \eprint{1610.00938}.

\bibitem[{\citenamefont{Vagnozzi}(2021)}]{Vagnozzi:2020gtf}
\bibinfo{author}{\bibfnamefont{S.}~\bibnamefont{Vagnozzi}}, \bibinfo{journal}{Mon. Not. Roy. Astron. Soc.} \textbf{\bibinfo{volume}{502}}, \bibinfo{pages}{L11} (\bibinfo{year}{2021}), \eprint{2009.13432}.

\bibitem[{\citenamefont{Benetti et~al.}(2022)\citenamefont{Benetti, Graef, and Vagnozzi}}]{Benetti:2021uea}
\bibinfo{author}{\bibfnamefont{M.}~\bibnamefont{Benetti}}, \bibinfo{author}{\bibfnamefont{L.~L.} \bibnamefont{Graef}}, \bibnamefont{and} \bibinfo{author}{\bibfnamefont{S.}~\bibnamefont{Vagnozzi}}, \bibinfo{journal}{Phys. Rev. D} \textbf{\bibinfo{volume}{105}}, \bibinfo{pages}{043520} (\bibinfo{year}{2022}), \eprint{2111.04758}.

\bibitem[{\citenamefont{Vagnozzi}(2023)}]{Vagnozzi:2023lwo}
\bibinfo{author}{\bibfnamefont{S.}~\bibnamefont{Vagnozzi}}, \bibinfo{journal}{JHEAp} \textbf{\bibinfo{volume}{39}}, \bibinfo{pages}{81} (\bibinfo{year}{2023}), \eprint{2306.16912}.

\bibitem[{\citenamefont{Zhu et~al.}(2023)\citenamefont{Zhu, Ye, and Cai}}]{Zhu:2023lbf}
\bibinfo{author}{\bibfnamefont{M.}~\bibnamefont{Zhu}}, \bibinfo{author}{\bibfnamefont{G.}~\bibnamefont{Ye}}, \bibnamefont{and} \bibinfo{author}{\bibfnamefont{Y.}~\bibnamefont{Cai}}, \bibinfo{journal}{Eur. Phys. J. C} \textbf{\bibinfo{volume}{83}}, \bibinfo{pages}{816} (\bibinfo{year}{2023}), \eprint{2307.16211}.

\bibitem[{\citenamefont{Papanikolaou et~al.}(2024)\citenamefont{Papanikolaou, Banerjee, Cai, Capozziello, and Saridakis}}]{Papanikolaou:2024fzf}
\bibinfo{author}{\bibfnamefont{T.}~\bibnamefont{Papanikolaou}}, \bibinfo{author}{\bibfnamefont{S.}~\bibnamefont{Banerjee}}, \bibinfo{author}{\bibfnamefont{Y.-F.} \bibnamefont{Cai}}, \bibinfo{author}{\bibfnamefont{S.}~\bibnamefont{Capozziello}}, \bibnamefont{and} \bibinfo{author}{\bibfnamefont{E.~N.} \bibnamefont{Saridakis}} (\bibinfo{year}{2024}), \eprint{2404.03779}.

\bibitem[{\citenamefont{Li et~al.}(2024)\citenamefont{Li, Lai, Xiang, and Wu}}]{Li:2024oru}
\bibinfo{author}{\bibfnamefont{C.}~\bibnamefont{Li}}, \bibinfo{author}{\bibfnamefont{J.}~\bibnamefont{Lai}}, \bibinfo{author}{\bibfnamefont{J.}~\bibnamefont{Xiang}}, \bibnamefont{and} \bibinfo{author}{\bibfnamefont{C.}~\bibnamefont{Wu}}, \bibinfo{journal}{JHEP} \textbf{\bibinfo{volume}{09}}, \bibinfo{pages}{138} (\bibinfo{year}{2024}), \eprint{2405.15889}.

\bibitem[{\citenamefont{Qiu and Zhu}(2024)}]{Qiu:2024sdd}
\bibinfo{author}{\bibfnamefont{T.}~\bibnamefont{Qiu}} \bibnamefont{and} \bibinfo{author}{\bibfnamefont{M.}~\bibnamefont{Zhu}} (\bibinfo{year}{2024}), \eprint{2408.06582}.

\bibitem[{\citenamefont{Lai and Li}(2025{\natexlab{b}})}]{Lai:2025efh}
\bibinfo{author}{\bibfnamefont{J.}~\bibnamefont{Lai}} \bibnamefont{and} \bibinfo{author}{\bibfnamefont{C.}~\bibnamefont{Li}} (\bibinfo{year}{2025}{\natexlab{b}}), \eprint{2504.19251}.

\bibitem[{\citenamefont{Cheung et~al.}(2016)\citenamefont{Cheung, Li, and Vergados}}]{Cheung:2016vze}
\bibinfo{author}{\bibfnamefont{Y.-K.~E.} \bibnamefont{Cheung}}, \bibinfo{author}{\bibfnamefont{C.}~\bibnamefont{Li}}, \bibnamefont{and} \bibinfo{author}{\bibfnamefont{J.~D.} \bibnamefont{Vergados}}, \bibinfo{journal}{Symmetry} \textbf{\bibinfo{volume}{8}}, \bibinfo{pages}{136} (\bibinfo{year}{2016}), \eprint{1611.04027}.

\bibitem[{\citenamefont{Boyle et~al.}(2004)\citenamefont{Boyle, Steinhardt, and Turok}}]{Boyle:2004gv}
\bibinfo{author}{\bibfnamefont{L.~A.} \bibnamefont{Boyle}}, \bibinfo{author}{\bibfnamefont{P.~J.} \bibnamefont{Steinhardt}}, \bibnamefont{and} \bibinfo{author}{\bibfnamefont{N.}~\bibnamefont{Turok}}, \bibinfo{journal}{Phys. Rev. D} \textbf{\bibinfo{volume}{70}}, \bibinfo{pages}{023504} (\bibinfo{year}{2004}), \eprint{hep-th/0403026}.

\bibitem[{\citenamefont{Cai}(2014)}]{Cai:2014bea}
\bibinfo{author}{\bibfnamefont{Y.-F.} \bibnamefont{Cai}}, \bibinfo{journal}{Sci. China Phys. Mech. Astron.} \textbf{\bibinfo{volume}{57}}, \bibinfo{pages}{1414} (\bibinfo{year}{2014}), \eprint{1405.1369}.

\bibitem[{\citenamefont{Li}(2024)}]{Li:2024dce}
\bibinfo{author}{\bibfnamefont{C.}~\bibnamefont{Li}}, \bibinfo{journal}{Phys. Rev. D} \textbf{\bibinfo{volume}{110}}, \bibinfo{pages}{083535} (\bibinfo{year}{2024}), \eprint{2407.10071}.

\bibitem[{\citenamefont{Li}(2025)}]{Li:2025ilc}
\bibinfo{author}{\bibfnamefont{C.}~\bibnamefont{Li}}, \bibinfo{journal}{Phys. Rev. D} \textbf{\bibinfo{volume}{111}}, \bibinfo{pages}{103517} (\bibinfo{year}{2025}), \eprint{2502.19124}.

\bibitem[{\citenamefont{Schmitz}(2021)}]{Schmitz:2020syl}
\bibinfo{author}{\bibfnamefont{K.}~\bibnamefont{Schmitz}}, \bibinfo{journal}{JHEP} \textbf{\bibinfo{volume}{01}}, \bibinfo{pages}{097} (\bibinfo{year}{2021}), \eprint{2002.04615}.

\bibitem[{\citenamefont{Annis et~al.}(2022)\citenamefont{Annis, Newman, and Slosar}}]{Annis:2022xgg}
\bibinfo{author}{\bibfnamefont{J.}~\bibnamefont{Annis}}, \bibinfo{author}{\bibfnamefont{J.~A.} \bibnamefont{Newman}}, \bibnamefont{and} \bibinfo{author}{\bibfnamefont{A.}~\bibnamefont{Slosar}} (\bibinfo{year}{2022}), \eprint{2209.08049}.

\bibitem[{\citenamefont{Bi et~al.}(2023)\citenamefont{Bi, Wu, Chen, and Huang}}]{Bi:2023tib}
\bibinfo{author}{\bibfnamefont{Y.-C.} \bibnamefont{Bi}}, \bibinfo{author}{\bibfnamefont{Y.-M.} \bibnamefont{Wu}}, \bibinfo{author}{\bibfnamefont{Z.-C.} \bibnamefont{Chen}}, \bibnamefont{and} \bibinfo{author}{\bibfnamefont{Q.-G.} \bibnamefont{Huang}}, \bibinfo{journal}{Sci. China Phys. Mech. Astron.} \textbf{\bibinfo{volume}{66}}, \bibinfo{pages}{120402} (\bibinfo{year}{2023}), \eprint{2307.00722}.

\bibitem[{\citenamefont{Sikivie}(1983)}]{Sikivie:1983ip}
\bibinfo{author}{\bibfnamefont{P.}~\bibnamefont{Sikivie}}, \bibinfo{journal}{Phys. Rev. Lett.} \textbf{\bibinfo{volume}{51}}, \bibinfo{pages}{1415} (\bibinfo{year}{1983}), \bibinfo{note}{[Erratum: Phys.Rev.Lett. 52, 695 (1984)]}.

\bibitem[{\citenamefont{Sikivie}(1985)}]{Sikivie:1985yu}
\bibinfo{author}{\bibfnamefont{P.}~\bibnamefont{Sikivie}}, \bibinfo{journal}{Phys. Rev. D} \textbf{\bibinfo{volume}{32}}, \bibinfo{pages}{2988} (\bibinfo{year}{1985}), \bibinfo{note}{[Erratum: Phys.Rev.D 36, 974 (1987)]}.

\bibitem[{\citenamefont{Sikivie et~al.}(2014)\citenamefont{Sikivie, Sullivan, and Tanner}}]{Sikivie:2013laa}
\bibinfo{author}{\bibfnamefont{P.}~\bibnamefont{Sikivie}}, \bibinfo{author}{\bibfnamefont{N.}~\bibnamefont{Sullivan}}, \bibnamefont{and} \bibinfo{author}{\bibfnamefont{D.~B.} \bibnamefont{Tanner}}, \bibinfo{journal}{Phys. Rev. Lett.} \textbf{\bibinfo{volume}{112}}, \bibinfo{pages}{131301} (\bibinfo{year}{2014}), \eprint{1310.8545}.

\bibitem[{\citenamefont{Chaudhuri et~al.}(2015)\citenamefont{Chaudhuri, Graham, Irwin, Mardon, Rajendran, and Zhao}}]{Chaudhuri:2014dla}
\bibinfo{author}{\bibfnamefont{S.}~\bibnamefont{Chaudhuri}}, \bibinfo{author}{\bibfnamefont{P.~W.} \bibnamefont{Graham}}, \bibinfo{author}{\bibfnamefont{K.}~\bibnamefont{Irwin}}, \bibinfo{author}{\bibfnamefont{J.}~\bibnamefont{Mardon}}, \bibinfo{author}{\bibfnamefont{S.}~\bibnamefont{Rajendran}}, \bibnamefont{and} \bibinfo{author}{\bibfnamefont{Y.}~\bibnamefont{Zhao}}, \bibinfo{journal}{Phys. Rev. D} \textbf{\bibinfo{volume}{92}}, \bibinfo{pages}{075012} (\bibinfo{year}{2015}), \eprint{1411.7382}.

\bibitem[{\citenamefont{Berlin et~al.}(2023)\citenamefont{Berlin, Blas, Tito~D'Agnolo, Ellis, Harnik, Kahn, Sch\"utte-Engel, and Wentzel}}]{Berlin:2023grv}
\bibinfo{author}{\bibfnamefont{A.}~\bibnamefont{Berlin}}, \bibinfo{author}{\bibfnamefont{D.}~\bibnamefont{Blas}}, \bibinfo{author}{\bibfnamefont{R.}~\bibnamefont{Tito~D'Agnolo}}, \bibinfo{author}{\bibfnamefont{S.~A.~R.} \bibnamefont{Ellis}}, \bibinfo{author}{\bibfnamefont{R.}~\bibnamefont{Harnik}}, \bibinfo{author}{\bibfnamefont{Y.}~\bibnamefont{Kahn}}, \bibinfo{author}{\bibfnamefont{J.}~\bibnamefont{Sch\"utte-Engel}}, \bibnamefont{and} \bibinfo{author}{\bibfnamefont{M.}~\bibnamefont{Wentzel}}, \bibinfo{journal}{Phys. Rev. D} \textbf{\bibinfo{volume}{108}}, \bibinfo{pages}{084058} (\bibinfo{year}{2023}), \eprint{2303.01518}.

\bibitem[{\citenamefont{Chen et~al.}(2025)\citenamefont{Chen, Li, Liu, Shu, Yang, and Zeng}}]{Chen:2023ryb}
\bibinfo{author}{\bibfnamefont{Y.}~\bibnamefont{Chen}}, \bibinfo{author}{\bibfnamefont{C.}~\bibnamefont{Li}}, \bibinfo{author}{\bibfnamefont{Y.}~\bibnamefont{Liu}}, \bibinfo{author}{\bibfnamefont{J.}~\bibnamefont{Shu}}, \bibinfo{author}{\bibfnamefont{Y.}~\bibnamefont{Yang}}, \bibnamefont{and} \bibinfo{author}{\bibfnamefont{Y.}~\bibnamefont{Zeng}}, \bibinfo{journal}{Rept. Prog. Phys.} \textbf{\bibinfo{volume}{88}}, \bibinfo{pages}{057601} (\bibinfo{year}{2025}), \eprint{2309.12387}.

\bibitem[{\citenamefont{Cai et~al.}(2010)\citenamefont{Cai, Saridakis, Setare, and Xia}}]{Cai:2009zp}
\bibinfo{author}{\bibfnamefont{Y.-F.} \bibnamefont{Cai}}, \bibinfo{author}{\bibfnamefont{E.~N.} \bibnamefont{Saridakis}}, \bibinfo{author}{\bibfnamefont{M.~R.} \bibnamefont{Setare}}, \bibnamefont{and} \bibinfo{author}{\bibfnamefont{J.-Q.} \bibnamefont{Xia}}, \bibinfo{journal}{Phys. Rept.} \textbf{\bibinfo{volume}{493}}, \bibinfo{pages}{1} (\bibinfo{year}{2010}), \eprint{0909.2776}.

\bibitem[{\citenamefont{Berlin et~al.}(2022)\citenamefont{Berlin, Blas, Tito~D'Agnolo, Ellis, Harnik, Kahn, and Sch\"utte-Engel}}]{Berlin:2021txa}
\bibinfo{author}{\bibfnamefont{A.}~\bibnamefont{Berlin}}, \bibinfo{author}{\bibfnamefont{D.}~\bibnamefont{Blas}}, \bibinfo{author}{\bibfnamefont{R.}~\bibnamefont{Tito~D'Agnolo}}, \bibinfo{author}{\bibfnamefont{S.~A.~R.} \bibnamefont{Ellis}}, \bibinfo{author}{\bibfnamefont{R.}~\bibnamefont{Harnik}}, \bibinfo{author}{\bibfnamefont{Y.}~\bibnamefont{Kahn}}, \bibnamefont{and} \bibinfo{author}{\bibfnamefont{J.}~\bibnamefont{Sch\"utte-Engel}}, \bibinfo{journal}{Phys. Rev. D} \textbf{\bibinfo{volume}{105}}, \bibinfo{pages}{116011} (\bibinfo{year}{2022}), \eprint{2112.11465}.

\bibitem[{\citenamefont{Domcke et~al.}(2022)\citenamefont{Domcke, Garcia-Cely, and Rodd}}]{Domcke:2022rgu}
\bibinfo{author}{\bibfnamefont{V.}~\bibnamefont{Domcke}}, \bibinfo{author}{\bibfnamefont{C.}~\bibnamefont{Garcia-Cely}}, \bibnamefont{and} \bibinfo{author}{\bibfnamefont{N.~L.} \bibnamefont{Rodd}}, \bibinfo{journal}{Phys. Rev. Lett.} \textbf{\bibinfo{volume}{129}}, \bibinfo{pages}{041101} (\bibinfo{year}{2022}), \eprint{2202.00695}.

\bibitem[{\citenamefont{Kahn et~al.}(2024)\citenamefont{Kahn, Sch\"utte-Engel, and Trickle}}]{Kahn:2023mrj}
\bibinfo{author}{\bibfnamefont{Y.}~\bibnamefont{Kahn}}, \bibinfo{author}{\bibfnamefont{J.}~\bibnamefont{Sch\"utte-Engel}}, \bibnamefont{and} \bibinfo{author}{\bibfnamefont{T.}~\bibnamefont{Trickle}}, \bibinfo{journal}{Phys. Rev. D} \textbf{\bibinfo{volume}{109}}, \bibinfo{pages}{096023} (\bibinfo{year}{2024}), \eprint{2311.17147}.

\bibitem[{\citenamefont{Fang et~al.}(2025)\citenamefont{Fang, Gao, Li, Shu, Wu, Xing, Xu, Xu, and Zhou}}]{Fang:2024ple}
\bibinfo{author}{\bibfnamefont{Y.}~\bibnamefont{Fang}}, \bibinfo{author}{\bibfnamefont{C.}~\bibnamefont{Gao}}, \bibinfo{author}{\bibfnamefont{Y.-Y.} \bibnamefont{Li}}, \bibinfo{author}{\bibfnamefont{J.}~\bibnamefont{Shu}}, \bibinfo{author}{\bibfnamefont{Y.}~\bibnamefont{Wu}}, \bibinfo{author}{\bibfnamefont{H.}~\bibnamefont{Xing}}, \bibinfo{author}{\bibfnamefont{B.}~\bibnamefont{Xu}}, \bibinfo{author}{\bibfnamefont{L.}~\bibnamefont{Xu}}, \bibnamefont{and} \bibinfo{author}{\bibfnamefont{C.}~\bibnamefont{Zhou}}, \bibinfo{journal}{Sci. China Phys. Mech. Astron.} \textbf{\bibinfo{volume}{68}}, \bibinfo{pages}{260301} (\bibinfo{year}{2025}), \eprint{2411.11294}.

\bibitem[{\citenamefont{Franciolini et~al.}(2022)\citenamefont{Franciolini, Maharana, and Muia}}]{Franciolini:2022htd}
\bibinfo{author}{\bibfnamefont{G.}~\bibnamefont{Franciolini}}, \bibinfo{author}{\bibfnamefont{A.}~\bibnamefont{Maharana}}, \bibnamefont{and} \bibinfo{author}{\bibfnamefont{F.}~\bibnamefont{Muia}}, \bibinfo{journal}{Phys. Rev. D} \textbf{\bibinfo{volume}{106}}, \bibinfo{pages}{103520} (\bibinfo{year}{2022}), \eprint{2205.02153}.

\bibitem[{\citenamefont{Gong et~al.}(2025)\citenamefont{Gong, Tian, Wu, and Zhu}}]{Gong:2025xsd}
\bibinfo{author}{\bibfnamefont{Y.}~\bibnamefont{Gong}}, \bibinfo{author}{\bibfnamefont{H.}~\bibnamefont{Tian}}, \bibinfo{author}{\bibfnamefont{L.}~\bibnamefont{Wu}}, \bibnamefont{and} \bibinfo{author}{\bibfnamefont{B.}~\bibnamefont{Zhu}} (\bibinfo{year}{2025}), \eprint{2504.09630}.

\bibitem[{\citenamefont{Gatti et~al.}(2024)\citenamefont{Gatti, Visinelli, and Zantedeschi}}]{Gatti:2024mde}
\bibinfo{author}{\bibfnamefont{C.}~\bibnamefont{Gatti}}, \bibinfo{author}{\bibfnamefont{L.}~\bibnamefont{Visinelli}}, \bibnamefont{and} \bibinfo{author}{\bibfnamefont{M.}~\bibnamefont{Zantedeschi}}, \bibinfo{journal}{Phys. Rev. D} \textbf{\bibinfo{volume}{110}}, \bibinfo{pages}{023018} (\bibinfo{year}{2024}), \eprint{2403.18610}.

\bibitem[{\citenamefont{Zeng et~al.}(2025)}]{SHANHE:2024tpr}
\bibinfo{author}{\bibfnamefont{Y.}~\bibnamefont{Zeng}} \bibnamefont{et~al.} (\bibinfo{collaboration}{SHANHE}), \bibinfo{journal}{Sci. Bull.} \textbf{\bibinfo{volume}{70}}, \bibinfo{pages}{661} (\bibinfo{year}{2025}), \eprint{2402.03432}.

\bibitem[{\citenamefont{Shu et~al.}(2024)\citenamefont{Shu, Xu, and Xu}}]{Shu:2024nmc}
\bibinfo{author}{\bibfnamefont{J.}~\bibnamefont{Shu}}, \bibinfo{author}{\bibfnamefont{B.}~\bibnamefont{Xu}}, \bibnamefont{and} \bibinfo{author}{\bibfnamefont{Y.}~\bibnamefont{Xu}} (\bibinfo{year}{2024}), \eprint{2410.22413}.

\bibitem[{\citenamefont{Arakawa et~al.}(2025)\citenamefont{Arakawa, Zaheer, Takhistov, Safronova, and Eby}}]{Arakawa:2025hcn}
\bibinfo{author}{\bibfnamefont{J.}~\bibnamefont{Arakawa}}, \bibinfo{author}{\bibfnamefont{M.~H.} \bibnamefont{Zaheer}}, \bibinfo{author}{\bibfnamefont{V.}~\bibnamefont{Takhistov}}, \bibinfo{author}{\bibfnamefont{M.~S.} \bibnamefont{Safronova}}, \bibnamefont{and} \bibinfo{author}{\bibfnamefont{J.}~\bibnamefont{Eby}} (\bibinfo{year}{2025}), \eprint{2502.08716}.

\bibitem[{\citenamefont{Fischer et~al.}(2024)}]{Fischer:2024nte}
\bibinfo{author}{\bibfnamefont{L.}~\bibnamefont{Fischer}} \bibnamefont{et~al.} (\bibinfo{year}{2024}), \eprint{2411.18346}.

\bibitem[{\citenamefont{Kohri et~al.}(2025)\citenamefont{Kohri, Terada, and Yanagida}}]{Kohri:2024qpd}
\bibinfo{author}{\bibfnamefont{K.}~\bibnamefont{Kohri}}, \bibinfo{author}{\bibfnamefont{T.}~\bibnamefont{Terada}}, \bibnamefont{and} \bibinfo{author}{\bibfnamefont{T.~T.} \bibnamefont{Yanagida}}, \bibinfo{journal}{Phys. Rev. D} \textbf{\bibinfo{volume}{111}}, \bibinfo{pages}{063543} (\bibinfo{year}{2025}), \eprint{2409.06365}.

\bibitem[{\citenamefont{Yin et~al.}(2025)\citenamefont{Yin, Kang, Jiao, and Rong}}]{Yin:2025jqe}
\bibinfo{author}{\bibfnamefont{Y.}~\bibnamefont{Yin}}, \bibinfo{author}{\bibfnamefont{R.}~\bibnamefont{Kang}}, \bibinfo{author}{\bibfnamefont{M.}~\bibnamefont{Jiao}}, \bibnamefont{and} \bibinfo{author}{\bibfnamefont{X.}~\bibnamefont{Rong}} (\bibinfo{year}{2025}), \eprint{2504.14944}.

\bibitem[{\citenamefont{Zhong et~al.}(2025)\citenamefont{Zhong, Fang, Zhan, Yang, and Huang}}]{Zhong:2025pto}
\bibinfo{author}{\bibfnamefont{H.}~\bibnamefont{Zhong}}, \bibinfo{author}{\bibfnamefont{Y.}~\bibnamefont{Fang}}, \bibinfo{author}{\bibfnamefont{J.}~\bibnamefont{Zhan}}, \bibinfo{author}{\bibfnamefont{R.}~\bibnamefont{Yang}}, \bibnamefont{and} \bibinfo{author}{\bibfnamefont{Y.}~\bibnamefont{Huang}}, \bibinfo{journal}{Eur. Phys. J. C} \textbf{\bibinfo{volume}{85}}, \bibinfo{pages}{436} (\bibinfo{year}{2025}).

\bibitem[{\citenamefont{Li et~al.}(2025{\natexlab{b}})\citenamefont{Li, Hong, and Zhang}}]{Li:2025eoo}
\bibinfo{author}{\bibfnamefont{J.-K.} \bibnamefont{Li}}, \bibinfo{author}{\bibfnamefont{W.}~\bibnamefont{Hong}}, \bibnamefont{and} \bibinfo{author}{\bibfnamefont{T.-J.} \bibnamefont{Zhang}}, \bibinfo{journal}{Astrophys. J.} \textbf{\bibinfo{volume}{985}}, \bibinfo{pages}{137} (\bibinfo{year}{2025}{\natexlab{b}}), \eprint{2504.13115}.

\end{thebibliography}
\bibliographystyle{apsrev}

\end{document}